\documentclass{aa}
\usepackage{graphicx}
\usepackage{txfonts}
\usepackage{natbib}
\usepackage{lscape}
\usepackage{aalongtable}

\begin{document}
   \title{Atomic diffusion and mixing in old stars}

   \subtitle{II. Observations of stars in the globular cluster NGC 6397 \\ with VLT/FLAMES-GIRAFFE \thanks{Based on data collected at European Southern Observatory (ESO), Paranal, Chile, under program ID 075.D-0125(A).}}

   \author{K. Lind\inst{1,2}\and
           A.\,J. Korn\inst{1}\and
           P.\,S. Barklem\inst{1}\and
           F. Grundahl\inst{3}
           }

   \institute{Department of Physics \& Astronomy, Uppsala University, Box 515, 751 20 Uppsala, Sweden\and
            European Southern Observatory (ESO), Karl-Schwarzschild-Strasse 2,
              857 48 Garching bei M\"unchen, Germany\\
              \email{klind@eso.org}\and
             Department of Physics \& Astronomy, University of Aarhus, Ny Munkegade, 8000 Aarhus C, Denmark
             }
   \date{Received 25/04/2008; accepted 25/08/2008}

  \abstract
   {Evolutionary trends in the surface abundances of heavier elements have recently been identified in the globular cluster NGC~6397 ([Fe/H]$=-2$), indicating the operation of atomic diffusion in these stars. Such trends constitute important constraints for the extent to which diffusion modifies the internal structure and surface abundances of solar-type, metal-poor stars.}
   {We perform an independent check of the reality and size of abundance variations within this metal-poor globular cluster.}
   {Observational data covering a large stellar sample, located between the cluster turn-off point and the base of the red giant branch, are homogeneously analysed. The spectroscopic data were obtained with the medium-high resolution spectrograph FLAMES/GIRAFFE on VLT-UT2 ($R\sim27\,000$). We derive independent effective-temperature scales from profile fitting of Balmer lines and by applying colour-$T_{\rm eff}$ calibrations to Str\"{o}mgren $uvby$ and broad-band $BVI$ photometry. An automated spectral analysis code is used together with a grid of MARCS model atmospheres to derive stellar surface abundances of Mg, Ca, Ti, and Fe.}
   {We identify systematically higher iron abundances for more evolved stars. The turn-off point stars are found to have $0.13$\,dex lower surface abundances of iron compared to the coolest, most evolved stars in our sample. There is a strong indication of a similar trend in magnesium,  whereas calcium and titanium abundances are more homogeneous. Within reasonable error limits, the obtained abundance trends are in agreement with the predictions of stellar structure models including diffusive processes (sedimentation, levitation), if additional turbulent mixing below the outer convection zone is included.}
   {}


   \maketitle

\section{Introduction}
A globular cluster (GC) constitutes, in many respects, a homogeneous stellar population and is therefore an excellent laboratory for testing and constraining models of stellar structure and evolution. Chemical abundances inferred from observations of GCs can give important clues to the physical processes at work in individual stars. Recently a discovery was made pointing to the existence of systematic differences in surface abundances between stars in different evolutionary stages in a metal-poor GC. \citet{Korn07} (hereafter Paper I) observe four samples of stars located between the main sequence (MS) turn-off (TO) point and the red giant branch (RGB) of NGC 6397, using FLAMES/UVES on the VLT, and found significant variations in abundances with effective temperature and surface gravity. In particular, they conclude iron to be under-abundant by $0.16\pm0.05$\,dex in the atmospheres of the TO stars compared to the RGB stars in their sample. Proposed as an explanation to this phenomenon is a continuous, long-term depletion of iron and other heavy elements from the surface layers, at work during the stars' life-time on the MS. The mechanism responsible for the depletion is generally referred to as atomic diffusion -- an umbrella term accounting for several diffusive processes. High ages and thin convective envelopes are factors that make metal-poor TO stars particularly prone to atomic diffusion. Once a star leaves the MS and begins to develop into the giant stage, its outer convection zone gradually reaches deeper layers, and elements that had previously been drained from the surface layers are mixed up again, eventually restoring the initial chemical composition (with fragile elements such as Li as notable exceptions). This is predicted to result in a gradual rise in the surface abundances of the depleted elements as the star evolves along the subgiant and red giant branches (hereafter, we write SGB for subgiant branch). GC stars presumably share the same original composition of elements such as iron, titanium, and calcium, which in combination with their high ages and low metal content make them suitable test cases for stellar structure models including atomic diffusion.\\
The abundance trend for iron presented in Paper I is directly contradictory to result by \citet{Gratton01}, who by studying TO vs.\ base-RGB (bRGB) stars conclude that there are no significant difference in iron abundance between the groups (magnesium is, however, found to be 0.15\,dex less abundant in the TO stars). The cause for the differing result is twofold: lower effective temperatures are found for the TO stars in Paper I and the analysis include stars on the RGB.\\
Here we present a follow-up analysis of spectroscopic data from FLAMES/GIRAFFE collected simultaneously to the observations presented in Paper I. A large sample of stars, $\approx 100$, covering the range from the TO to the bRGB is homogeneously analysed, with the aim of further constraining possible variations in surface abundance in the cluster. In particular, we investigate whether the results presented in Paper I are robust for a larger sample of stars, analysed with an independent set of tools. We start by describing the observations and the data-reduction procedure in Sect.\ 2. Sect.\ 3 is devoted to the determination of fundamental stellar parameters and a description of the spectrum analysis code. In Sect.\ 4 we present and discuss our abundance results. Sect.\ 5 compares predictions from stellar-structure models including atomic diffusion to the abundances we have inferred from observations. In Sect.\ 6 we summarise our conclusions.\\

\section{Observations and data reduction}
\subsection{Photometric observations}
\begin{figure}
       \centering
       \includegraphics[width=7cm,viewport=1cm 0cm 8cm 9.5cm]{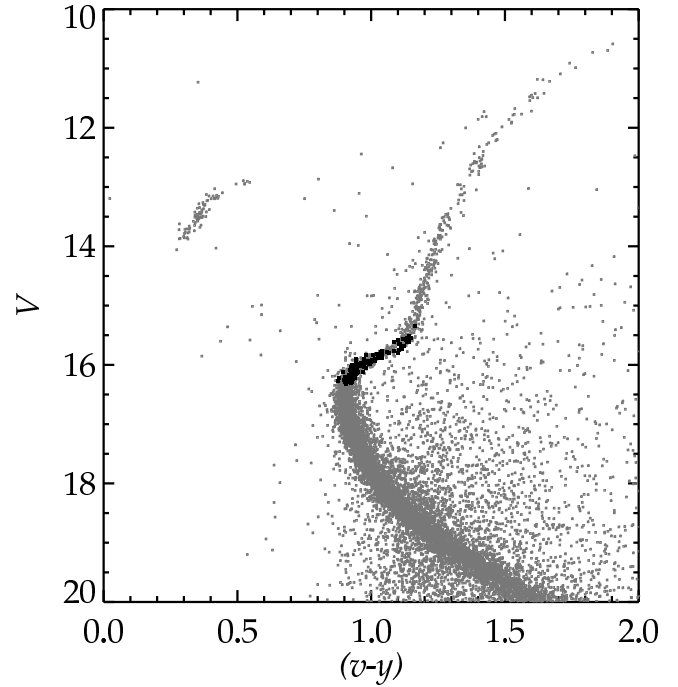}
       \caption{Observed colour-magnitude diagram of NGC 6397. The targets selected for the spectroscopic study are marked with black squares.}
       \label{fig:cmd}
\end{figure}

The target selection for the spectroscopic study is based on Str\"{o}mgren $uvby$ photometry. The photometric observations were collected with the DFOSC instrument on the 1.5\,m telescope on La Silla, Chile, in 1997. Additional $BVI$ photometric data were obtained in 2005. The UCAC2 catalog at Vizier \citep{Ochsenbein00} provides astrometry with a precision of $\sim0\farcs10$ for individual stars. A full description of the photometric observations and their reduction is given in Paper I and references therein. \\ 
We select a sample of 135 stars for spectroscopic analysis, located between the cluster TO point at $V=16.3$ to the bRGB at $V=15.2$. Apart from providing the target list, the photometry is also considered in this work in the determination of stellar parameters. Fig.\ \ref{fig:cmd} shows a colour-magnitude diagram, $V$ versus $(v-y)$, of the cluster, with the selected stars marked.  \\

\subsection{Spectroscopic observations}
All spectroscopic data were collected in Service Mode, with the fibre-fed, multi-object, medium-high resolution spectrograph FLAMES/GIRAFFE \citep{Pasquini02} at ESO-VLT. FLAMES allows for 132 objects to be observed simultaneously, with GIRAFFE in MEDUSA mode. A total exposure time of 20.5\,h was spent under grey-to-bright lunar conditions, but with good seeing (average seeing $\sim$\,0.7$\arcsec$). We apply four of the highest resolution 'B-settings', which together cover a total spectral range of 822\,{\AA} in the optical. On average 15 fibres per observation are dedicated to a simultaneous monitoring of the sky background to ensure a proper sky-correction. Stars in different evolutionary stages are distributed randomly across the FLAMES field-of-view. We find no gradient in background light over the field and subtract a single, averaged sky-spectrum from the stellar spectra obtained during each observation. The setup is briefly summarised in Table \ref{tab:setup}.

\begin{table}
      \caption{Observational setup for FLAMES/GIRAFFE $^a$.}
         \label{tab:setup}
         \centering
         \begin{tabular}{ccccr}
                \hline\hline
                Setting     &   Wavelength     & Resolution  &  Exposure & Dates 2005 \\
                \#            &       range [{\AA}]  &     $\lambda / \Delta \lambda$ & time&  \\
                \hline
                5B            &       4\,376 - 4\,552      & 26\,000  & 7h\,40m   & 23,26-29\,Mar \\
                7B            &       4\,742 - 4\,932      & 26\,700  & 5h\,10m   & 29-31\,Mar \\
                9B            &       5\,143 - 5\,356      & 25\,900  & 3h\,50m   & 01-03\,Apr \\
                14B           &       6\,383 - 6\,626      & 28\,800  & 3h\,50m   & 02-04\,Apr \\
                \hline
                & & & \\
                \multicolumn{5}{l}{$^a$ Data from http://www.eso.org/instruments/flames}\\
         \end{tabular}
\end{table}

\noindent Basic reduction of the spectroscopic data is performed with the ESO-maintained GIRAFFE pipeline, version 1.0, using standard settings. Further processing of the data is carried out with MIDAS and C-routines (kindly provided by N. Christlieb). The GIRAFFE CCD chip has a defect in its upper right corner caused by the illumination of a diode, giving elevated dark-current values to a significant number of pixels. If uncorrected for, this will propagate into a flux upturn towards longer wavelengths in dozens of spectra. The feature can be removed by subtracting a dark frame from the raw science frames (after proper scaling with exposure time). To avoid the introduction of additional noise we chose to correct for dark current after the spectra had been extracted, when the glow is a well-behaved and easily smoothable function of wavelength. For this purpose we pseudo-reduce a dark frame produced close in time to the observations. We find, however, that it is necessary to apply an additional, minor scaling factor to the dark counts to completely remove the feature, which we ascribe to the documented time-variability of the glow strength\footnote{see the instrument's quality control pages:\\ http://www.eso.org/observing/dfo/quality/GIRAFFE/qc/dark\_qc1.html}.\\
The radial velocity correction of our stellar spectra results in a cluster mean heliocentric radial velocity of $18.1\pm0.3\rm\,km\,s^{-1}$ from measurements of 133 stars. We find a velocity dispersion of $3.40\rm\,km\,s^{-1}$. This radial velocity value is in agreement with the recent result by \citet{Milone06} who derive $18.36\pm 0.09 (\pm 0.10)\rm\,km\,s^{-1}$ (the first is the random error and the second the systematic error), obtained from 1486 stars. We identify two stars that are not members of the cluster, 502219 and 501856, both with deviating radial velocities and evidently too metal-rich spectra (see Sec. 4). \\
After sky-subtraction and radial velocity correction the spectra are coadded and rebinned onto a final wavelength scale with a step size of 0.1\,{\AA}. The $S/N$-ratio of the coadded spectra in the 14B setting varies from 28 to 148 per rebinned pixel with a mean value of 85, estimated from a relatively line-free region between 6\,440\,-6\,449\,{\AA}. The corresponding mean value for the bluer 7B setting is 76, from measurements between $4\,812-4\,821$\,{\AA}.


\section{Analysis}

In this section we describe how we derive the fundamental parameters necessary to conduct the abundance analysis. We emphasise that we have made an effort to carry out all steps of the investigation using independent tools from those of Paper I, in particular by implementing different spectral synthesis and model atmospheres codes. 

\begin{figure}
       \centering
       \includegraphics[width=6.5cm, angle=90]{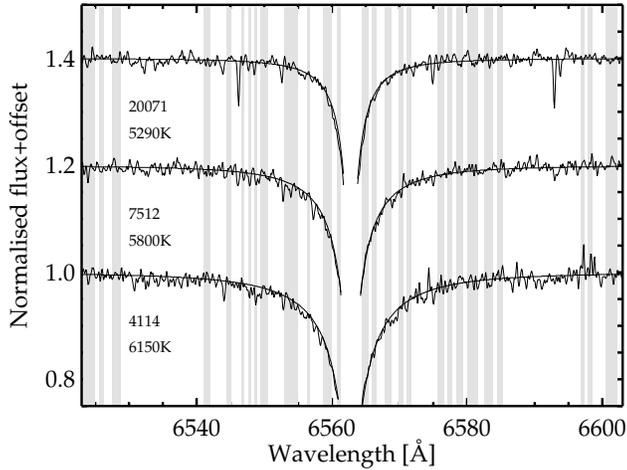}
       \caption{The best-fit H$\alpha$ profiles for three stars. The thick line corresponds to a model with the stated effective temperature. The grey-shaded regions represent the regions used for fitting.}
       \label{fig:halpha}
\end{figure}

\begin{figure}
       \centering
       \includegraphics[width=9cm,viewport= 1.5cm 1cm 21cm 25cm]{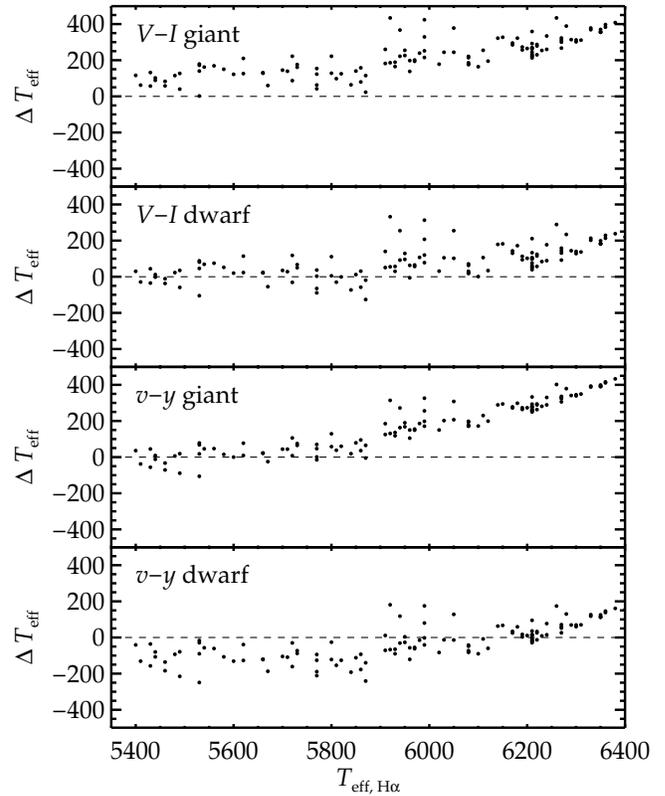}
       \caption{Comparison between our H$\alpha$-based effective temperatures and those obtained by applying the calibrated relations of \citet{Alonso96,Alonso99} to colour indices $v-y$ and $V-I$. The y-axis shows $\Delta T_{\rm eff}=T_{\rm eff, H\alpha}-T_{\rm photometry}$ for the stated index and calibration.}
       \label{fig:temp}
\end{figure}

\subsection{Effective temperatures}
\begin{figure}
       \centering
       \includegraphics[width=6.5cm,angle=90]{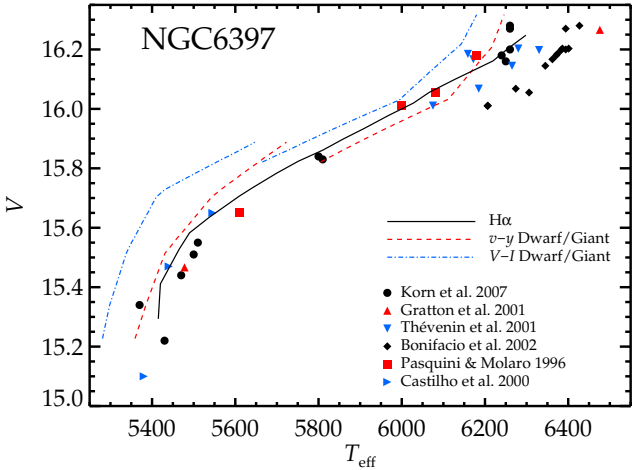}
       \caption{Comparison between our spectroscopic $T_{\rm eff}$-scale, the photometric values obtained from $v-y$ and $V-I$, and the results from six other studies. }
       \label{fig:other}
\end{figure}

Profile fitting of Balmer lines is a well-established $T_{\rm eff}$-indicator in solar-type stars. Considering the reliable flat-fielding which can be obtained with fibre-fed spectroscopy and the large free spectral range of GIRAFFE (FSR\,$\sim$\,200\,{\AA}), this data set is well suited for analysis of broad lines such as the Balmer lines. We implement an automated $\chi^2$-minimization technique for the normalisation and profile fitting of the wings of H$\alpha$. Synthetic spectra are produced with the FORTRAN code HLINPROF \citep{Barklem03}\footnote{Available at www.astro.uu.se/$\sim$barklem.}, which is incorporated in the LTE (local thermodynamical equilibrium) spectrum synthesis code SYNTH \citep{Piskunov92}. We treat Stark broadening according to the tabulated line profiles of \citet{Stehle99} and self broadening (resonance broadening and van der Waals broadening) according to \citet{Barklem00}. Radiative broadening and an estimate of line broadening due to helium collisions are also included in the calculations. \\
To compute a set of plane-parallel model atmospheres for the spectrum synthesis we use the MARCS code \citep{Gustafsson75,Asplund97}. MARCS is a one-dimensional (1D), LTE model atmospheres code, invoking flux conservation (radiative and convective) and hydrostatic equilibrium. Convection is treated according to the customary mixing length theory, with the mixing length parameters set to $\alpha=1.5$ and $y=0.076$. \\
The observed spectrum is compared to a grid of synthetic spectra, equidistant in effective temperature with 10\,K steps. The fitting only considers regions that are not significantly affected by absorption from metal and telluric lines (see Fig. \ref{fig:halpha}) as judged from high-resolution and high-$S/N$ spectra of metal-poor subgiants. First, an initial guess of the correct model is made and only wavelength regions located $\pm35-50$\AA\ from the H$\alpha$-line center are used. Here the sensitivity of the hydrogen line to effective temperature is low and the difference between observed and modelled flux is minimised to find the optimal straight line used for normalisation. Thereafter, only wavelength regions in the effective-temperature sensitive parts are used, located within $\pm25$\AA\ of the center, to find the model that minimises the $\chi^2$-distance to the observed flux. However, to avoid modelling uncertainties affecting the line core, only residual flux above 0.75 is considered. The whole procedure is iterated until the initial guess coincides with the resulting $T_{\rm eff}$-value. \\
By repeatedly adding random noise to synthetic spectra and feeding them as input to the procedure, the precision of the fitting method can be determined. We find that a typical $S/N$-ratio of 85 gives a random error of about 30\,K, for the coolest as well as for the hottest stars. For our lowest-quality spectra the corresponding value is 80\,K and for the highest quality data the random error is as low as 15\,K. The absolute error in $T_{\rm eff}$ is more difficult to assess, as it depends on several factors in addition to observational uncertainty, of which the most important ones are the line broadening of hydrogen, the treatment of convection in the stellar atmosphere, and the stellar parameters $\log{g}$ and [Fe/H]. \citet{Barklem02} show that these contributions together may add up to an error of the order of 100\,K for stars in this metallicity and effective-temperature range.\\
From the FLAMES solar atlas\footnote{http://www.eso.org/observing/dfo/quality/GIRAFFE/pipeline/\\solar.html} we retrieve reduced GIRAFFE spectra from observations illuminating all MEDUSA fibers in the 14B-setting, which we average into a single solar spectrum. Using the setup described above we derive $T_{\rm eff,H\alpha}=5\,630$\,K for the Sun, which is approximately 150\,K lower than the accepted value, an offset which we ascribe mainly to observational uncertainty. \citet{Barklem02} derive a solar effective temperature of $5\,722\pm81\rm\,K$ with the same models, using an H$\alpha$-spectrum from the Kitt Peak atlas \citep{Kurucz84}. The H$\alpha$ line profile found from the FLAMES/GIRAFFE solar atlas is more narrow compared to e.g. the Kitt peak atlas and we recommend to re-observe the Sun with GIRAFFE with the new CCD (upgraded in May 2008) and seek the reason for this difference. \\ 
Using our obtained solar effective temperature as a zero-point offset, we consequently shift all effective temperatures by +150\,K. We thus assume that, in the stellar-parameter range spanned by our targets, the H$\alpha$-line has a similar sensitivity to effective temperature as the Sun, which may be realistic to a first order approximation. The shifted H$\alpha$-based effective temperatures are the ones we later adopt in the abundance analysis and we will refer to them as the spectroscopic $T_{\rm eff}$ in the remaining part of the paper. \\
To assess the validity of our spectroscopic effective temperatures we construct photometric $T_{\rm eff}$ scales from four different colour indices. The \citet{Alonso96,Alonso99} relations, calibrated on the infrared flux method, are used to derive effective temperatures based on $b-y$, $v-y$, $B-V$, and $V-I$. Note that, at this metallicity and stellar-parameter space, the corresponding relations of \citet{Ramirez05} produce very similar effective temperatures (see Paper I). There are two Alonso et al.\ calibrations for each index, one suitable for main sequence stars of spectral types F0-K5 \citep{Alonso96} and one for giant stars in the same spectral range \citep{Alonso99}\footnote{Calibrations for the $v-y$-index for dwarf and giant stars are published in Paper I.}. Since our targets range from TO stars at the very end of the MS to bRGB stars, both the dwarf and giant calibrations are used to separately derive effective temperatures. Note that by doing so, we also apply the $T_{\rm eff}-\rm colour$ relations to stars that do not fall in the range in colour covered by the calibrations at this metallicity, which may affect the results. \\
Before applying the calibrations, we eliminate star-to-star scatter caused by errors in the observed colours. We clean the photometric sample based on data quality, as measured by the DAPHOT $SHARP$ parameter, and construct fiducial relations between each colour index and the $V$ magnitude, which is the best observed and calibrated quantity. The relations are obtained by averaging the colours for stars in $V$ bins of 0.22 magnitudes. We then shift the observed colour for each star onto the sequence and, finally, we correct all colours for the reddening of NGC 6397. This value has been measured in several studies and typical estimates lie in the range $E(B-V)=0.17-0.20$ (see e.g.\ \citeauthor{Reid98} 1998). Here a value of $0.179$ is adopted, following \citet{AnthonyTwarog00}. The reddening in the other colour indices are derived from $E(B-V)$, using the relation coefficients given in \citet{Ramirez05} and $E(v-y)=1.7\times E(b-y)$.\\
Fig.\ \ref{fig:temp} shows a comparison between our derived spectroscopic and photometric effective temperatures for 122 stars (all stars for which the H$\alpha$ line is observed). The difference between the H$\alpha$-based values and the effective temperatures obtained from the dwarf and giant calibrations of the narrow-band index $v-y$ and the broad-band index $V-I$ are plotted against $T_{\rm eff, H\alpha}$. The agreement between H$\alpha$ and the $v-y$ dwarf calibration scale is good for the hotter half of the sample, with spectroscopic $T_{\rm eff} \geq 5\,900$\,K. The average difference for individual stars is here $\sim 70\,\rm K$. For the cooler half, at spectroscopic $T_{\rm eff} \leq 5\,900$\,K, good agreement is instead seen with the giant calibration, with a mean difference $\sim50\rm\,K$. The spectroscopic temperature scale hence suggests a transition between the two $v-y$ calibration scales at some point in the middle of the SGB, in line with what may be expected. The other narrow-band index, $b-y$ (not shown), produces very similar effective temperatures to the $v-y$ scales. The $V-I$ dwarf calibration scale appears similar in relative behaviour to $v-y$, but the effective temperatures are shifted towards slightly cooler values, giving a good agreement with the spectroscopic $T_{\rm eff}$ for the cooler half of the $T_{\rm eff}$-range. The giant scale, however, is evidently too cool overall to reproduce the spectroscopic results. For $B-V$ (not shown) the two Alonso et al. calibrations result in scales that are identical to each other to within 5\,K. In comparison to the H$\alpha$-based values the agreement is reasonable at the hotter end whereas the cooler end is offset to higher temperatures by up to $\sim200$\,K. This index thus implies a markedly shorter total effective temperature range for our sample than the other three.\\
As an additional test, we also compared the effective temperatures obtained with the Alonso et al. relations to $(v-y)$ and $(b-y)$-calibrations based on synthetic MARCS colours (\citeauthor{Gustafsson08} 2008; calibrations to be published in \"{O}nehag et al. 2008, in preparation). The theoretical calibrations reproduce the $T_{\rm eff}$-range spanned by the targets well, indicating a somewhat shorter total range, by about $\sim70\,\rm K$, than what is predicted from a combination of the two Alonso et al.\ relations for $b-y$ and $v-y$, respectively.\\
Much effort has previously been made to pin down the effective-temperature scale of this cluster at the SGB. A summary of the results from six high-resolution studies, including Paper I, is given in Fig.\ \ref{fig:other}, along with our spectroscopic and four photometric scales. In the plot, which illustrates visual magnitude versus effective temperature, filled symbols represent previous results, and lines represent this study. For clarity, we show only a smoothed relation between $V$ and $T_{\rm eff,H\alpha}$ (the $1\sigma$-scatter around this line is 90\,K) and only parts of the photometric $V-T_{\rm eff}$ relations based on $v-y$ and $V-I$, i.e.\ the \citet{Alonso96} calibration is applied only to the hotter half and the \citet{Alonso99} only to the cooler half of the stellar sample. All data are collected from the stated publications or from references therein. Only the effective temperatures that are actually adopted by the different authors are shown, even if several different scales are presented in the study. The visual magnitudes in the sample of \citet{Thevenin01} are from their Table 1, taking the average whenever two values are given. \\
Fig.\ \ref{fig:other} shows that the different studies agree within approximately $200-300\,\rm K$ at a given visual magnitude, which may be a reasonable error, albeit undesirably large when the aim is precise chemical abundances. Most studies focus on TO stars and the only two studies, except the present, that include stars on both ends of the SGB are Paper I and \citet{Gratton01}. On a relative scale, which is most essential for this work, we predict a slightly larger effective temperature range between $V=15.5-16.2$ than Paper I and a shorter range than Gratton et al. In the former study, effective temperatures are also obtained by H$\alpha$ profile fitting but using a different model atmosphere and spectrum synthesis code (ODF-MAFAGS, \citeauthor{Grupp04I} 2004, \citeauthor{Fuhrmann93} 1993). For the eight targets in common with this study we find group-averaged effective temperatures that are different by less than 10\,K. The typical difference for individual stars is 50\,K. The $T_{\rm eff}$ value found for the TO-star group of the Gratton et al.\ study exceeds our estimate by almost $200\,\rm K$ and is extensively discussed in Paper I, where the authors point out that the effective-temperature assignment of Gratton et al.\,, based on H$\alpha$, may be systematically affected by the echelle-order blaze residuals imprinted on the continuum in the UVES (slit-mode) observations. Differences in model atmospheres and hydrogen line-broadening theory used may also explain part of the offset (see Paper I).

\subsection{Surface gravities}
We derive surface gravities for the target stars using the customary relation between effective temperature, luminosity, mass, and surface gravity. Luminosities are calculated from the apparent visual magnitude $V$ and the \citet{Alonso99} calibration for bolometric correction, which is given as a function of metallicity, [Fe/H]\footnote{${\rm[X/Y]}=\log{\left(\frac{N_{\rm X}}{N_{\rm Y}}\right)}-\log{\left(\frac{N_{\rm X}}{N_{\rm Y}}\right)}_{\sun}$ where $N_{\rm X}$ is the number density of element X.}, and $T_{\rm eff}$. The metallicity is set to $-2.0$ for all stars. The distance modulus of NGC 6397 is assumed to be 12.57. Stellar masses are inferred from a 13.5\,Gyr isochrone of the cluster \citep{Richard05}, which places the stars in the mass range $0.78\rm\,M_{\sun}-0.79\rm\,M_{\sun}$. With these values, the surface gravity found for TO stars at $T_{\rm eff}=6\,250\,$K is $\log{g}=3.96$ and $\log{g}=3.40$ for bRGB stars at $T_{\rm eff}=5\,450\,$K. We use a single, averaged value of $\log{g}$ for all stars for which we have derived the same $T_{\rm eff}$. \\
The aim of this study is to draw conclusions about abundance differences between stars. We are therefore mainly interested in the accuracy with which we can determine surface gravities on the relative, rather than the absolute scale. In this respect, the effective temperatures have the largest, albeit small, influence on the surface gravity values. A rise in effective temperature of 100\,K corresponds to an increase in logarithmic surface gravity of approximately 0.03\,dex. This can be compared to an increase in stellar mass by $0.01\rm\,M_{\sun}$ that propagates into a rise in $\log{g}$ by 0.005\,dex. We set a constant metallicity, to avoid circular arguments as regards the existence of abundance trends. The influence of [Fe/H] on the derived surface gravities is anyway small. An increase in [Fe/H] by 0.1\,dex propagates into a decrease in $\log{g}$ of 0.006\,dex. We estimate the relative error in $V$ to approximately 0.01 magnitudes, which gives a contribution to the error in $\log{g}$ of 0.004\,dex.
The value of the distance modulus only affects the absolute values of the surface gravities. An increase in $(m-M)_{V}$ of 0.5 corresponds to an overall rise in $\log{g}$ of 0.2\,dex. Given these low dependencies of relative photometric surface-gravity values on other parameters, abundance differences within the sample can be constrained from gravity-sensitive lines (e.g.\ those of Fe\,II) with high confidence.

\begin{figure}
       \centering
       \includegraphics[width=6.5cm,angle=90]{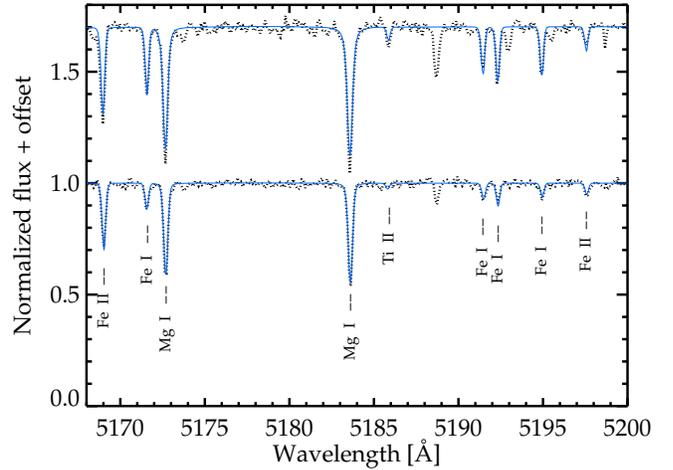}
       \caption{A selected part of two observed, normalised spectra (dotted lines) and the best fitting model spectra (solid lines), as determined by the analysis code. Note that the models in this figure only contain the spectral lines that are used in the abundance analysis. These lines are marked with labels. The top spectrum belongs to the bRGB star 23267 and the bottom plot to the TO star 12318}
       \label{fig:spec}
\end{figure}

\subsection{Spectrum analysis code}
For abundance analysis of the stellar sample we use the same automated code as developed for the Hamburg/ESO R-process enhanced star survey, (HERES, \citeauthor{Barklem05} 2005). We summarise here some important aspects of the analysis code, referring the reader to \citeauthor{Barklem05} for a full description. \\
The software is built on the IDL and C++ based Spectroscopy Made Easy (SME) package by \citet{Valenti96}, with subsequent updates, but has been modified to run without any user interaction. Some improvements have also been made to the original SME spectrum synthesis code, most importantly the inclusion of continuous scattering in the source function. As the software is implemented, we supply it with 1) a count spectrum and the measurement error in each pixel, 2) the final stellar parameters $T_{\rm eff}$ and $\log{g}$ and an initial guess for the metallicity ([Fe/H]$=-2.0$), 3) a list of lines to base the abundance analysis on and wavelength windows where each line can be found and 4) a grid of MARCS \citep{Asplund97} plane-parallel model atmospheres, all with scaled solar abundances except for an enhancement of alpha-elements of 0.4\,dex. This setup is the same as implemented in \citet{Barklem05}, except that we keep $\log{g}$ fixed.\\
The code automatically normalises a spectral region stretching 7\,{\AA} on both sides of each given wavelength window, by iterative fitting of a low-order polynomial. At this point any remaining cosmic ray hits are also identified and discarded. The line central wavelength is determined and, if necessary, shifted slightly (within the estimated uncertainty) to coincide with the model wavelength. Observed and modelled spectra are then compared and the best statistical match is determined via a parameter-optimization algorithm \citep{Marquardt63,Press92}. We assume the stars to be slow rotators and fix the projected rotational velocity to $1\rm\,km\,s^{-1}$. The free parameters are thus the abundance, the microturbulence $\xi$, and the macroturbulence $v_{\rm macro}$ (Gaussian, isotropic), in which also the instrumental broadening is included. The online appendix contains a list of all lines used in the abundance analysis, with references to the $gf$-values that were adopted. Based on the availability of lines covered in the spectra, we choose to derive abundances of iron, titanium, calcium, and magnesium. Fig.\ \ref{fig:spec} shows a selected part of two observed, normalised spectra and the best-fit model spectra.\\
The analysis assumes LTE line formation in one-dimensional, plane-parallel, model atmospheres. These simplifying assumptions can, of course, influence the derived abundances to a certain extent, which we discuss further in Sect.\ 4. In particular, strong line cores are expected to be poorly described by LTE photospheric models. To avoid large biases, pixel values below 50\,\% of the continuum flux are generally disregarded. \\
To analyse the resulting abundances, we also need knowledge about how errors in the stellar parameters affect them. By rerunning the code with one perturbed parameter at a time, holding the others fixed, the abundance sensitivity to these quantities are found for the elements considered. We make three such additional runs, with $T_{\rm eff}+100$\,K, $\log{g}+0.3$\,dex, and $\xi+0.2\rm\,km\,s^{-1}$.\\


\section{Results}

\begin{figure*}
       \centering
       \includegraphics[width=12cm,viewport=2cm 8cm 20cm 19cm,angle=90]{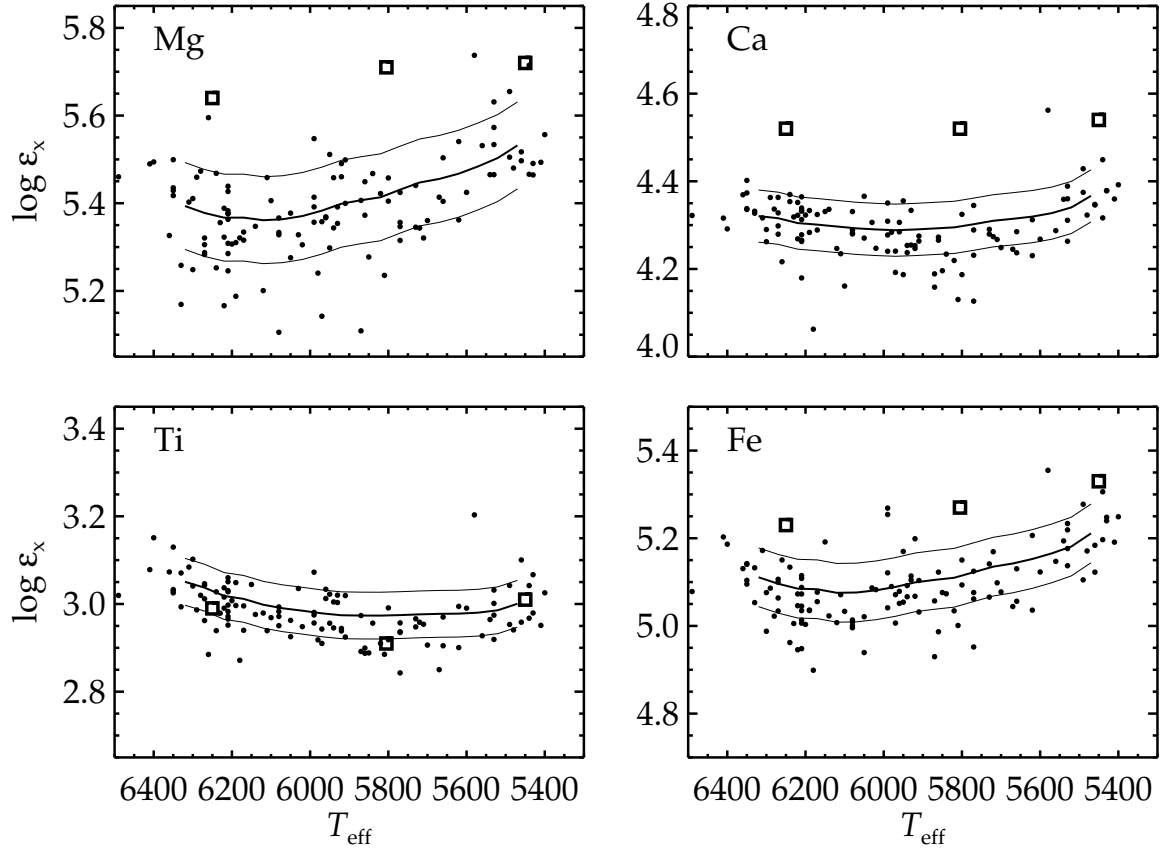}
       \caption{Abundance trends (filled dots) with effective temperature for Mg, Ca, Ti, and Fe from analysis of 116 stars. The thick lines display general trends and the thin lines show estimated 1$\sigma$ limits. Open squares represent the results from Paper I, in which Mg, Ca, and Fe are treated in NLTE and Ti in LTE.}
       \label{fig:abund}
\end{figure*}

Fig.\ \ref{fig:abund} shows our derived chemical abundances of magnesium, calcium, and iron, for 116 stars. Six stars, for which we have only two observations each, were rejected due to largely deviating abundance results (not shown), most likely caused by the low quality of these spectra. General trends are obtained by averaging the abundances in effective-temperature bins of 500\,K (with shrinking box-size towards each end) and then applying additional smoothing to avoid the influence from outliers. The thin lines shown in the plot indicate the standard deviations of the trends' residuals. For comparison, Fig.\ \ref{fig:abund} also shows the results from Paper I (open squares). Note that the abundances of Fe, Mg, and Ca then were derived by non-LTE (hereafter NLTE) analysis.\\
It is evident that our derived abundances indeed suggest that there are trends with effective temperature on the SGB. Especially in iron there is a notable trend of increased abundances towards lower $T_{\rm eff}$. The magnesium values display a similar behaviour, but due to larger star-to-star scatter the trend is less clearly visible. The calcium abundances may point to a slight, insignificant, increase towards cooler effective temperature, whereas the titanium abundances rather indicate the opposite, equally insignificant, trend with $T_{\rm eff}$. Table \ref{tab:abund} summarises the average abundances we derive for all elements at two representative effective temperature points, 5\,450\,K and 6\,250\,K, and the estimated 1$\sigma$ scatter in $\log\epsilon$\footnote{$\log{\epsilon_{\rm x}}=\log{\left(\frac{N_{\rm x}}{N_{\rm H}}\right)}+12$}. The microturbulence values are found to gradually decrease from a value $1.86\pm0.13\rm\,km\,s^{-1}$ for the stars at $T_{\rm eff}=6\,250$\,K to $1.47\pm0.09\rm\,km\,s^{-1}$ for the stars at $5\,450$\,K. \\
A preliminary analysis of the stars found to be non-members based on radial-velocity measurements (see Sect. 2) reveals that they are both subgiants, with approximate stellar parameters $T_{\rm eff}=5\,700$\,K and $\log{g}=3.3$. We find $\rm[Fe/H]=-0.6$ for 501856 and $\rm[Fe/H]=-1.1$ for 502219, both indicating $\rm[Ca/Fe]=0.3$. Among the cluster stars, 7720 is found to have consistently outlying abundances. With $T_{\rm eff}=5\,580$\,K, it displays abundances exceeding the average by $0.19-0.26$\,dex. The star has excellent agreement between H$\alpha$-based and photometry-based effective temperatures and a radial velocity typical of the cluster. \\

\begin{figure}
       \centering
       \includegraphics[viewport=3.5cm 0.2cm 13cm 9.5cm, width=6.5cm]{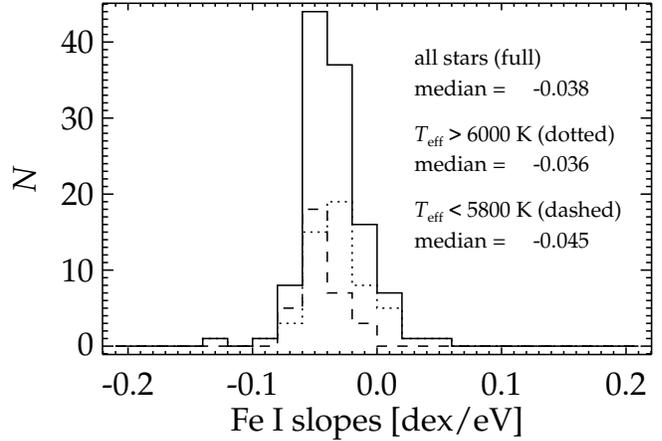}
       \caption{A histogram of the slopes obtained for linear fits of iron abundance with excitation potential of Fe\,I lines.}
       \label{fig:ex}
\end{figure}

\begin{table*}
      \caption{Average abundances obtained at two effective temperature points. }
         \label{tab:abund}
         \centering
         \begin{tabular}{cccccccccccc}
                \hline\hline
                &$T_{\rm eff}$ [K]&  $\log{g}$ [cgs] & $\xi$ [$\rm km\,s^{-1}$] & $\log{\epsilon_{\rm Mg}}$ & $\sigma_{\rm Mg}$ & $\log{\epsilon_{\rm Ca}}$ &
                $\sigma_{\rm Ca}$  & $\log{\epsilon_{\rm Ti}}$         & $\sigma_{\rm Ti}$ & $\log{\epsilon_{\rm Fe}}$ &
                $\sigma_{\rm Fe}$\\
                \hline
                         & 6\,250  &  3.96 & 1.86 & 5.37   & 0.10 & 4.31    & 0.06 & 3.03 & 0.05 & 5.09     & 0.07\\
                         & 5\,450  &  3.40 & 1.47 & 5.54   & 0.10 & 4.38    & 0.06 & 3.00 & 0.05 & 5.22     & 0.07\\
                \hline
                $\Delta$ & 800     &  0.56 & 0.39 & $-$0.17&      & $-$0.07 &      & 0.03 &      & $-$0.13  &     \\
                \hline
         \end{tabular}
\end{table*}

\subsection{Error sources}
\begin{figure}
       \centering
       \includegraphics[width=6.5cm,angle=90]{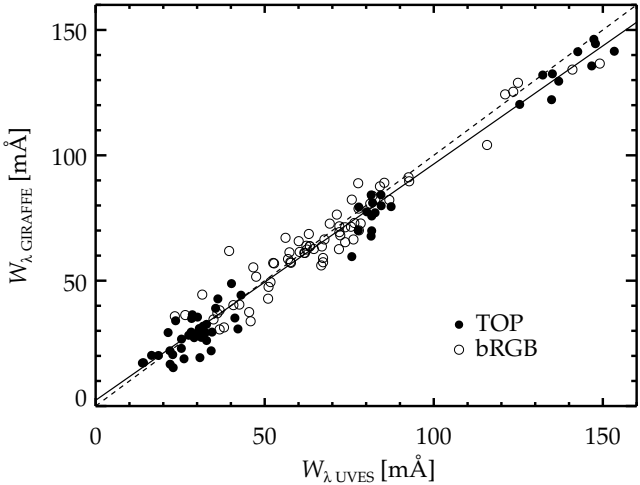}
       \caption{The figure shows a comparison between equivalent widths measured for the same lines in spectra of the same stars, obtained with GIRAFFE and UVES, respectively. The filled circles corresponds to stars located at the TO and the open circles to bRGB stars. The dashed line mark 1:1 agreement and the solid line is a linear fit to the measurements.}
       \label{fig:eqwi}
\end{figure}

There are several possible sources of error to consider, which may effect our abundances on the relative and absolute scale. First of all our analysis relies on the assumption of LTE line formation, which is not in general a good approximation for metal-poor stars. Particularly lines from neutral minority species are believed to form out of LTE. NLTE corrections are expected to be larger in cool metal-poor stars than in their metal-rich counterparts, mainly because the reduced amount of metals gives less line-blocking in the UV, which causes more over-ionisation. Also, the amount of free electrons available in the atmosphere is reduced in metal-poor stars, which means less LTE-establishing electron collisions. Considering the narrow range in stellar parameter space covered by our targets, we would expect NLTE corrections to be similar for all stars, thereby mainly influencing the abundances on the absolute scale.\\
The second point we consider is the abundance dependence on stellar parameters, most importantly on effective temperature. Table \ref{tab:sens} shows what effect perturbing $T_{\rm eff},\log{g}$, and $\xi$ from their derived values has on the derived abundances of all elements, at two effective temperature points. With this knowledge it is easily predicted how systematic errors in e.g.\ our $T_{\rm eff}$ scale would influence the results. When discussing this issue for each element, we take into account that changes in the effective temperature scale cause changes in surface gravity (see Sect.\ 3.2). We neglect possible changes in $\xi$ caused by altering $T_{\rm eff}$ or $\log{g}$. \\
The most influential parameter on the chemical abundances is the effective temperature. As seen in Sec.\ 3.1, $T_{\rm eff}$-determinations based on H$\alpha$ and cluster photometry seem to be in good agreement, but additional information can be drawn from the excitation equilibria of the targets. Generally, the abundance of each element is the one that gives the optimal match to all lines, but for testing purposes we also compute individual abundances for all neutral iron lines and thus can compare our adopted temperature scale with that implied by excitation equilibrium. The resulting slopes obtained for iron abundance with excitation potential of the lines are shown in a histogram in Figure \ref{fig:ex}. The median slope for the sample is negative, $-0.038\rm\,dex/eV$, i.e.\ overall cooler effective temperatures are needed to establish the equilibria. By estimate, the equilibria of the hottest stars, $T_{\rm eff}>6000\rm\,K$, are reproduced with a lowering of the effective temperatures of approximately 270\,K. The stars in the cooler half, $T_{\rm eff}<5800\rm\,K$, require a similar change, 260\,K, as the abundance slopes with excitation potential are more negative than the hotter half, but also have a higher temperature-sensitivity. There is thus no sign of a sizeable expansion or contraction of the adopted effective-temperature scale.\\
A third source of error originates in the spectrum itself. We may judge how well GIRAFFE performs in comparison to the higher resolution spectrograph UVES as eight of our observed stars, five TO stars and three bRGB stars, are the same targets as presented in Paper I. The results from measurements of a number of equivalent widths of lines in the GIRAFFE and UVES spectra of these stars are displayed in Fig.\ \ref{fig:eqwi}. The equivalent widths measured in the GIRAFFE spectra are plotted against the corresponding value obtained from the same lines in the UVES spectra. Filled circles represent TO stars and open circles bRGB stars. The results indicate that the GIRAFFE spectra have slightly weaker lines compared to those obtained with UVES. The difference amounts to approximately 4\,\%, which in logarithmic abundance units corresponds to $-0.02$\,dex (assuming a linear relation between abundance and equivalent width). There is no obvious difference between TO and bRGB stars, leading us to conclude that this has no impact on the relative abundances we derive for our sample stars.  \\

\subsection{Magnesium}
Our magnesium abundances are based on two of the Mg\,Ib triplet lines, 5\,172\,{\AA} and 5\,183\,{\AA}. As seen in Table \ref{tab:sens}, magnesium is the most sensitive of all elements to $T_{\rm eff}$ and $\log{g}$, which may contribute to explaining the higher level of scatter, approximately 0.10\,dex, seen in $\log{\epsilon}_{\rm Mg}$ compared to the other elements. Another probable source of scatter is conversion of Mg to Al, which is observed as an inverse correlation between the two elements in GC stars (see e.g.\ \citeauthor{Gratton04} 2004). \\
As seen in Table \ref{tab:abund} we obtain a difference of 0.17\,dex between stars at the cooler and hotter end of the SGB. For a trend of this size to vanish, i.e.\, for our magnesium abundances to agree over the whole sample, the effective temperature of the TO stars would have to be raised by at least 320\,K, and more, around 450\,K, if one takes into consideration a simultaneous rise in surface gravity of 0.03\,dex per 100\,K (See Sect.\ 3.2), since the two stellar parameters have opposite influence on the magnesium abundance. Alternatively, the effective temperature of the bRGB stars would have to be lowered by $170-220$\,K. We note that the trend we obtain of 0.17\,dex is larger than the 0.08\,dex found in Paper I for the difference in magnesium abundance between their bRGB stars at average $T_{\rm eff}=5\,456\,$K and TO stars at $T_{\rm eff}=6\,254\,$K. \\
Based on existing studies, NLTE corrections for lines of neutral magnesium are positive in these stars, due to over-ionization. Paper I, following \citet{Gehren04}, calculate $\log{\epsilon}_{\rm NLTE}-\log{\epsilon}_{\rm LTE}$ to range between $0.11$\,dex and $0.14$\,dex in their sample, based on Mg I 5\,528\,{\AA}. Hence there are only small differential effects in abundance, judging from this line. \citet{Mashonkina08} find an average NLTE correction of 0.12\,dex for the somewhat hotter, but otherwise very similar, metal-poor star HD 84937, from five Mg\,I lines, including 5\,528\,{\AA}, 5\,172\,{\AA}, and 5\,183\,{\AA}. \\
NLTE effects are pronounced in the line cores of the triplet lines, in this case leading to the observed lines being deeper than the LTE-based prediction. These influences are partly avoided for the coolest stars in the sample, for which the line cores extend deeper than 50\% of the continuum flux and are disregarded by the spectrum analysis code. In the hottest stars, all pixels contributing to the lines are evaluated. Given this methodology, differential effects on the abundance trend obtained from photospheric NLTE modelling, including the line core, cannot be excluded. \\

\subsection{Calcium}
Nine lines of neutral calcium are considered in the determination of $\log{\epsilon}_{\rm Ca}$. The strongest lines, i.e.\ those consisting of the highest number of pixels and thereby having the largest influence on the derived abundances, are 4\,434\,{\AA}, 4\,454\,{\AA}, 6\,462\,{\AA}, and 6\,439\,{\AA}.
Our average calcium abundances increase from 4.31\,dex at the TO to 4.38\,dex at the bRGB, i.e.\ a trend of 0.07\,dex. This can be compared to 0.02\,dex found in Paper I, based on 6\,122\,{\AA}, 6\,162\,{\AA}, and 6\,439\,{\AA}. Table \ref{tab:sens} shows that $\log{\epsilon_{\rm Ca}}$ has a low sensitivity to changes in $T_{\rm eff}$ and $\log{g}$ and the scatter in abundance is also small. To completely remove a trend of 0.07\,dex in calcium abundance, one would have to raise the TO star effective temperatures by $500-750$\,K (assuming the dependence is linear also for such a large perturbation), again depending on the corresponding change in surface gravity, or equivalently lower the bRGB star effective temperatures by $210-260$\,K. \\
Following \citet{Mashonkina07} we estimate the NLTE corrections to the calcium abundances for the 6\,439{\AA} and 4\,454{\AA} lines. For the former line, we find a minor abundance correction of +0.05\,dex for the TO stars and no correction at all for the bRGB stars, while the latter line points to +0.13\,dex at the hotter end and +0.10\,dex at the cooler end. We would therefore expect that the trend in calcium abundance resulting from NLTE analysis of these lines is almost flat. Considering the possibility of differential NLTE abundance corrections of this size, it is clear that our inferred trend as such is insignificant. \\

\subsection{Titanium}
The titanium abundances are based on twelve Ti\,II lines and three Ti\,I lines, all except two (Ti\,II 5\,188\,{\AA} and 5\,226\,{\AA}) situated in the spectral range $4\,390-4\,540$\,{\AA}. LTE line formation is believed to be a fair approximation for Ti\,II, since this is the dominant ionization stage. For neutral titanium one would on the other hand not expect LTE to be valid, but no thorough NLTE analysis exists in the literature covering this wavelength range for metal-poor stars. In any case, singly ionised titanium lines dominate neutral ones in our abundance analysis, both in terms of number and line strengths, and we therefore assume NLTE corrections to be of lesser importance.\\
On the absolute scale, the Ti abundances are in good agreement with the results from Paper I, which are based on Ti\,II $ 5\,188$\,{\AA} and 5\,226\,{\AA}. We obtain abundances ranging from 3.03\,dex at 6\,250\,K to 3.00\,dex at 5\,450\,K, and the abundances from Paper I fall within our 1$\sigma$ limit of 0.04\,dex from these values. The abundance difference we find between TO stars and bRGB stars is $-0.03$\,dex, whereas Paper I report a positive trend of 0.02\,dex. \\
Due to the large influence from Ti\,II lines, $\log{\epsilon_{\rm Ti}}$ has a positive reaction to a rise in surface gravity contrary to the other elements and is also highly insensitive to changes in effective temperature at the hotter end. Lowering the surface gravity of the TO stars by 0.15\,dex, or alternatively raising the bRGB surface gravity by $0.21$\,dex, would result in a flat trend. 

\subsection{Iron}
We include 51 Fe\,I lines and 9 Fe\,II lines in our analysis. Considering that the latter are comparatively weak, we estimate that lines from Fe\,I bear higher weight. We find a trend of 0.13\,dex between TO and bRGB stars, again slightly larger than the 0.10\,dex reported in Paper I. The average iron abundance of the sample ranges from 5.09\,dex to 5.22\,dex, corresponding to $\rm[Fe/H]=-2.41$ to $-2.28$, using $\log{\epsilon}_{\rm Fe,\sun}=7.5$. To flatten the trend, one would need to raise the TO star $T_{\rm eff}$ by $350-390$\,K or lower $T_{\rm eff}$ at the bRGB by $190-230$\,K. 
Following \citet{Korn03}, Paper I find a minor NLTE abundance correction for neutral iron at the level of 0.03-0.05\,dex in TO stars as well as in RGB stars and practically no differential NLTE effects between the groups. The small size of the corrections is likely due to the adoption of high rates of collisions with neutral hydrogen, and we do  not exclude that NLTE corrections may adjust our iron abundances significantly upward. \\
Neglecting differential NLTE influences and other possible modelling deficits, the significance level of the identified trend in iron abundance depends only on errors in relative stellar parameters. An estimate of the error in $\Delta\log{\epsilon_{\rm Fe}}$ between TO and bRGB stars can be obtained by adding the individual errors listed in Table 3 in quadrature for the hottest and coolest stars, respectively. Assuming we have constrained the range of the effective-temperature scale to better than 100\,K, the surface-gravity scale within 0.1\,dex, and the microturbulence scale within $0.1\rm\,km\,s^{-1}$, the total error in abundance is 0.050\,dex at $T_{\rm eff}=6250\rm\,K$ and 0.080\,dex at $T_{\rm eff}=5450\rm\,K$. A lower limit on the significance of the trend is thus $0.13/0.080=1.6$\,sigma, corresponding to approximately 90\%.\\

\subsection{Comparison to other studies}
In general our results for the TO stars agree well with the LTE abundances obtained for the sample included in the study by \citet{Thevenin01}, whilst their NLTE values are significantly higher. Taking the average of their LTE values (given in their  Table 4) results in abundances that lie up to 0.1\,dex above the findings in this work. Similar difference is seen between our iron abundances and the LTE values obtained in Paper I. There is a larger offset, $\sim$0.15\,dex, between the LTE abundances for magnesium and calcium. We estimate that the small differences in stellar parameters between the studies can account for up to 0.1\,dex in logarithmic abundance and that the remaining part originates in line selection and modelling techniques. This is confirmed by a re-analysis of two of the UVES targets in Paper I, using the same automated procedure and line list as we have implemented in this study. However, for the discussion of abundance trends, differences in absolute abundance on this level are irrelevant. \\
Other values reported from medium- or high-resolution spectroscopic studies of NGC 6397 are $\rm[Fe/H]=-2.03\pm0.02\pm0.04$ \citep{Gratton01}, $\rm[Fe/H]=-2.0\pm0.05$ \citep{Castilho00}, $\rm[Fe/H]=-2.21\pm0.05$ \citep{Gratton82}, $-1.88$ \citep{Gratton89}, $-1.99\pm0.01$ \citep{Minneti93}, $-1.82\pm0.04$ \citep{Caretta97}. We note that several of the mentioned studies are performed exclusively on giant stars.   
 
\begin{table}
      \caption{Sensitivity of logarithmic abundances to stellar parameters.}
         \label{tab:sens}
         \centering
         \begin{tabular}{ccccc}
                \hline\hline
                Element         &   $T_{\rm eff}$ &     $T_{\rm eff}$+100\,K     &  $\log{g}+0.1$\,dex  &   $\xi+0.1\rm\,km\,s^{-1}$ \\
                \hline
             Mg      &  6\,250      &    0.053 &  -0.053 & -0.048    \\
             Mg      &  5\,450      &    0.097 &  -0.075 & -0.025    \\
             Ca      &  6\,250      &    0.014 &  -0.016 & -0.025    \\
             Ca      &  5\,450      &    0.033 &  -0.023 & -0.032    \\
             Ti      &  6\,250      &    0.001 &   0.019 & -0.029    \\
             Ti      &  5\,450      &    0.014 &   0.014 & -0.036    \\
             Fe      &  6\,250      &    0.037 &  -0.014 & -0.031    \\
             Fe      &  5\,450      &    0.066 &  -0.025 & -0.038    \\
                \hline
         \end{tabular}
\end{table}


\begin{figure}
       \centering
       \includegraphics[width=12cm,viewport=1cm 1.5cm 20cm 14cm,angle=90]{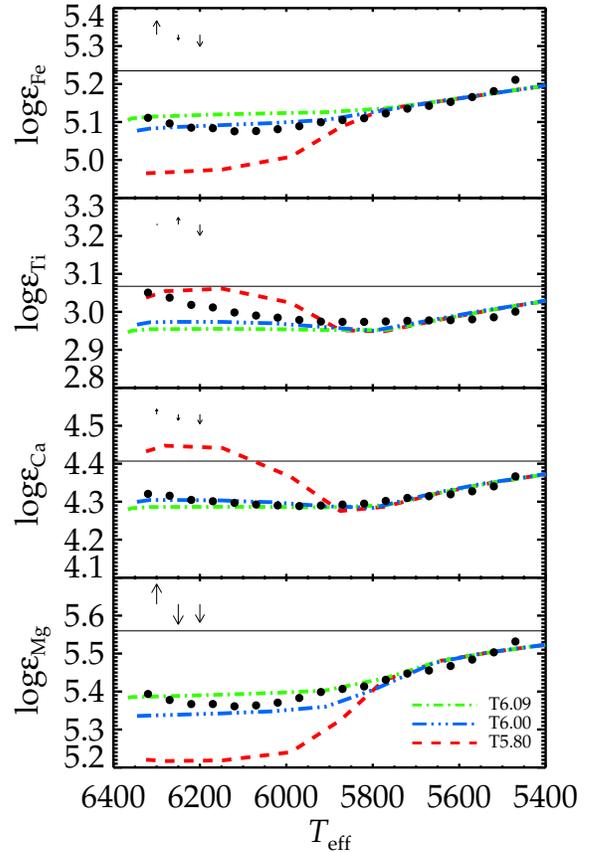}
       \caption{Comparison between the abundance trends from Fig.\ \ref{fig:abund} (here displayed with black dots) and the predictions from stellar structure models including atomic diffusion and turbulent mixing. The horizontal, black, solid lines represent the initial abundances of the models. The three arrows in the upper left corner of each plot indicate the abundance change at $T_{\rm eff}=6\,250\,K$ when raising the effective temperature by 100\,K (left), raising the surface gravity by 0.1\,dex (middle), and raising the microturbulence values by 0.1\,km/s (right).}
       \label{fig:model}
\end{figure}

\section{Discussion}
Abundance variations in GCs of the kind our analysis suggests may be explained with stellar-structure models that allow for radial diffusion of chemical elements. We compare our obtained abundances to predictions from models of Population II stars by \citet{Richard02I,Richard05}, which account for all the physics of particle transport that can be modelled from first principles (see Paper I, Richard et al. 2005, and references therein). \\
The main free parameter in these models is the introduction of turbulent transport of unknown physical nature, which the authors conclude to be best parameterised as a (decreasing) function of temperature and density. Additional mixing is needed to meet the strong observational evidence of a flat and thin plateau of Li in metal-poor TO and SGB stars over a wide range in metallicities, known as the 'Spite plateau' of lithium \citep{Spite82}. Omitting turbulent mixing, the plateau cannot be reproduced by the models, as too much Li diffuses into the stellar interior. In Fig.\ \ref{fig:model} we illustrate the predictions from three models implementing different values of a reference temperature $T_0$, which controls the efficiency of the mixing. The models are denoted by T5.80 (lowest efficiency), T6.00 (the best fitting efficiency to the results in Paper I), and T6.09 (highest efficiency) and represent the range in least efficient mixing that is compatible with the Spite plateau of lithium. The absolute abundances predicted by the three models are slightly adjusted, individually for each element, to agree with our derived abundances in the cooler half of the effective-temperature range ($\leq5800$\,K), where the three models coincide. Here, the region mixed by turbulence is fully encompassed by the convective envelope and the mixing efficiency is irrelevant for the surface abundances. The black solid horizontal lines mark the initial abundances of the models (accounting for the vertical shifts), $\log{\epsilon}_{\rm Mg}=5.56$, $\log{\epsilon}_{\rm Ca}=4.41$, $\log{\epsilon}_{\rm Ti}=3.07$, and $\log{\epsilon}_{\rm Fe}=5.23$, or $\rm[Fe/H]=-2.27$.  \\
The comparison shows that the predicted increase in surface abundances from $T_{\rm eff}\approx5\,800$\,K to 5\,400\,K is overall well reproduced by our obtained values. We further conclude that, within reasonable error bars, the best-fitting model found in Paper I (T6.00) is also a good choice for this sample of stars. Generally, it is clear that the surface iron and magnesium abundances predicted from models including atomic diffusion and mixing are more compatible with the results of our analysis than flat trends are. \\

\section{Conclusions}
We have presented a homogeneous analysis of medium-high resolution spectroscopic data covering a large sample of stars, located between the TO point and bRGB of the globular cluster NGC 6397. The obtained iron abundances show a significant trend with evolutionary stage. The magnesium abundances also indicate the presence of a trend of similar size, but the high sensitivity of this element to stellar parameters combined with the possibility of differential NLTE corrections prevent an unambiguous detection. The calcium and titanium abundances can be reconciled with a flat behaviour. The difference in iron abundances between the stars at $T_{\rm eff,H\alpha}=5\,450\,$K and stars at $T_{\rm eff,H\alpha}=6\,250\,$K amounts to $0.13$\,dex in $\log{\epsilon_{\rm Fe}}$, a trend that is robust to realistic errors in stellar parameters. Raising the effective temperatures of the hottest stars by $\sim 350$\,K would remove the trend in iron abundance. Considering that the total range in effective temperature between the considered points is $800$\,K, this would imply an error in our $T_{\rm eff}$-scale of $43\,$\%, which we regard as unlikely. Alternatively, the same effect may be achieved by lowering the effective temperatures of the coolest stars by $\sim200$\,K, correspoding to a systematic error of almost 25\,\%. Even that is an improbably large correction, given the good agreement between spectroscopy, through profile-fitting of H$\alpha$ and the different photometric calibrations that were exploited. The excitation equilibrium of Fe\,I points to cooler $T_{\rm\,eff}$-values overall, but on the relative scale the agreement is satisfactory. Overall, the results of this study independently support the conclusions of \citet{Korn07}, indicating that atomic diffusion significantly affects the surface abundances of metal-poor stars near the main turn-off point.\\
The abundance analysis is based on traditional LTE analysis using 1D, hydrostatic model atmospheres. Although it seems that corrections to the obtained abundances of Fe and Mg by 1D, NLTE modelling have no large differential impact in comparison to the sizes of the found trends, we cannot rule out that our results are artifacts from insufficient modelling techniques. More sophisticated models, implementing NLTE line formation in 3D, dynamical model atmospheres may provide a more definite answer. Some explorations with LTE line formation using 3D models were performed in Paper I, indicating that the results are robust in this respect. \\     
The obtained abundance trends with effective temperature are, for magnesium, calcium, and iron, well reproduced by stellar-structure models including atomic diffusion and turbulent mixing. Especially, the good relative agreement seen at the cooler half of the SGB, where additional turbulent transport is irrelevant, indeed lends support to the notion that atomic diffusion shapes the surface abundances of unevolved metal-poor stars.
 
\begin{acknowledgements}
We thank O.\,Richard for providing stellar-structure models and A.\"{O}nehag for letting us apply a yet unpublished colour-$T_{\rm eff}$ calibration. B.\,Gustafsson is thanked for valuable discussions and comments to the work. We thank the referee who helped improve the paper. K.\,L. is grateful to F.\,Primas and J.\,Sundqvist for their support and proof-readings of the manuscript. A.\,J.\,K.\ acknowledges research fellowships by the Leopoldina Foundation/Germany (under grant BMBF-LPD 9901/8-87). F.\,G.\ acknowledges support from the Danish AsteroSeismology Centre, the Carlsberg Foundation, and the Instrument Center for Danish Astronomy (IDA). P.\,B.\ is a Royal Swedish Academy of Sciences Research Fellow supported by a grant from the Knut and Alice Wallenberg Foundation. A.\,J.\,K.\ and P.\,B.\ also acknowledge support from the Swedish Research Council. This research has made use of the VizieR catalogue access tool, CDS, Strasbourg, France.
\end{acknowledgements}

\nocite{Bonifacio02}
\nocite{Pasquini96}
\bibliographystyle{aa}
\bibliography{0051refs}
\Online
\begin{landscape}
\small
\begin{table}
\caption{Stellar parameters and abundances derived for the sample stars. Targets marked with * are the targets of Korn et al. (2007). }
\begin{tabular}{lllllllllllllllll}
\hline\hline 
ID  & $\alpha\rm(2000)$ & $\delta\rm(2000)$ &$V$& $T_{\rm H\alpha}$ & $\log{g}$ & $\xi$ & $\log{\epsilon_{\rm Mg}}$ & $\log{\epsilon_{\rm Ca}}$  & $\log{\epsilon_{\rm Ti}}$ & $\log{\epsilon_{\rm Fe}}$  & $v-y$ & $T_{v-y,\rm Dwarf}$ &$T_{v-y,\rm Giant}$ & $V-I$& $T_{V-I, \rm Dwarf}$& $T_{V-I, \rm Giant}$ \\ 
 &  & & & [K] &[cgs] & $\rm[km\,s^{-1}]$ & & & & & & [K]& [K] & & [K] & [K] \\
\hline 
10174   &17 40 10.680 &53 37 41.30 & 16.171   &    6220 &  3.93 &  1.86 &  5.39 & 4.35 & 3.02 & 5.07  &   0.693 & 6194 & 5931 &  0.583  &6105 & 5941    \\
10197*  &17 40 10.770 &53 38 26.40 & 16.160   &    6190 &  3.91 &  1.98 &  5.19 & 4.28 & 3.05 & 5.04  &   0.695 & 6187 & 5926 &  0.585  &6096 & 5933    \\
10268   &17 40 11.160 &53 40 26.20 & 16.163   &    6200 &  3.92 &  1.83 &  5.31 & 4.32 & 3.01 & 5.00  &   0.694 & 6189 & 5927 &  0.585  &6098 & 5935    \\
10372   &17 40 11.680 &53 44 30.40 & 15.543   &    5410 &  3.31 &  1.48 &  5.49 & 4.36 & 2.95 & 5.19  &   0.892 & 5541 & 5448 &  0.761  &5439 & 5348    \\
10387   &17 40 11.480 &53 38 45.60 & 16.113   &    6080 &  3.85 &  1.78 &  5.33 & 4.28 & 2.95 & 5.01  &   0.702 & 6159 & 5907 &  0.594  &6058 & 5902    \\
10672   &17 40 12.460 &53 40 41.60 & 15.587   &    5480 &  3.47 &  1.53 &  5.48 & 4.32 & 2.94 & 5.17  &   0.881 & 5573 & 5474 &  0.755  &5457 & 5365    \\
10676   &17 40 12.610 &53 43 41.10 & 16.317   &    6400 &  4.06 &  1.98 &  5.49 & 4.29 & 3.15 & 5.19  &   0.680 & 6246 & 5965 &  0.566  &6180 & 6005    \\
10902   &17 40 13.340 &53 42 39.30 & 16.060   &    6080 &  3.85 &  1.90 &  5.11 & 4.33 & 2.99 & 5.00  &   0.710 & 6128 & 5886 &  0.603  &6017 & 5866    \\
10917   &17 40 13.250 &53 39 53.80 & 15.865   &    5770 &  3.67 &  1.60 &  5.35 & 4.29 & 2.94 & 5.12  &   0.785 & 5864 & 5700 &  0.675  &5733 & 5616    \\
10976   &17 40 13.710 &53 45 36.40 & 15.804   &    5660 &  3.61 &  1.68 &  5.50 & 4.29 & 2.97 & 5.13  &   0.810 & 5782 & 5639 &  0.701  &5639 & 5532    \\
11066   &17 40 13.710 &53 39 15.40 & 16.238   &    6350 &  4.02 &  1.90 &  5.42 & 4.37 & 3.03 & 5.14  &   0.685 & 6225 & 5951 &  0.573  &6151 & 5980    \\
11153   &17 40 13.980 &53 39 33.40 & 16.312   &    6270 &  3.97 &  1.88 &  5.31 & 4.30 & 3.01 & 5.06  &   0.680 & 6244 & 5964 &  0.567  &6178 & 6003    \\
11318   &17 40 14.500 &53 38 56.90 & 15.683   &    5460 &  3.44 &  1.33 &  5.50 & 4.35 & 3.10 & 5.12  &   0.856 & 5644 & 5531 &  0.743  &5497 & 5402    \\
11794   &17 40 15.840 &53 34 36.30 & 15.940   &    5940 &  3.77 &  1.76 &  5.34 & 4.24 & 2.95 & 5.07  &   0.753 & 5971 & 5777 &  0.645  &5848 & 5718    \\
12082   &17 40 16.740 &53 35 11.70 & 15.802   &    5660 &  3.61 &  1.56 &  5.40 & 4.24 & 2.90 & 5.06  &   0.811 & 5780 & 5637 &  0.702  &5636 & 5529    \\
12318*  &17 40 17.640 &53 39 34.20 & 16.182   &    6210 &  3.93 &  1.73 &  5.38 & 4.31 & 3.00 & 5.11  &   0.691 & 6201 & 5935 &  0.581  &6113 & 5948    \\
12387   &17 40 18.090 &53 44 57.50 & 16.173   &    6210 &  3.93 &  1.73 &  5.38 & 4.36 & 3.03 & 5.09  &   0.693 & 6195 & 5931 &  0.583  &6106 & 5942    \\
12473   &17 40 18.340 &53 44 38.50 & 16.097   &    6280 &  3.97 &  1.92 &  5.47 & 4.34 & 3.02 & 5.02  &   0.704 & 6150 & 5901 &  0.597  &6046 & 5891    \\
12881   &17 40 19.640 &53 45 23.30 & 15.980   &    5990 &  3.80 &  1.81 &  5.36 & 4.31 & 2.96 & 5.12  &   0.736 & 6031 & 5819 &  0.629  &5912 & 5775    \\
13053   &17 40 19.900 &53 39 49.00 & 16.010   &    6140 &  3.88 &  1.83 &  5.35 & 4.34 & 2.98 & 5.02  &   0.724 & 6077 & 5851 &  0.617  &5961 & 5818    \\
13093   &17 40 20.100 &53 41 47.10 & 16.187   &    6330 &  4.01 &  1.87 &  5.17 & 4.33 & 3.07 & 5.13  &   0.690 & 6204 & 5937 &  0.580  &6117 & 5952    \\
13160   &17 40 20.350 &53 43 18.40 & 15.821   &    5700 &  3.63 &  1.55 &  5.36 & 4.25 & 2.91 & 5.08  &   0.803 & 5805 & 5656 &  0.694  &5665 & 5555    \\
13466   &17 40 21.040 &53 40 01.20 & 15.829   &    5990 &  3.80 &  1.77 &  5.41 & 4.35 & 3.07 & 5.27  &   0.800 & 5815 & 5664 &  0.690  &5677 & 5566    \\
13535   &17 40 21.380 &53 43 05.30 & 15.988   &    6030 &  3.82 &  1.69 &  5.33 & 4.31 & 3.04 & 5.09  &   0.733 & 6043 & 5828 &  0.626  &5925 & 5786    \\
13552   &17 40 21.360 &53 41 43.40 & 16.043   &    6110 &  3.87 &  1.66 &  5.46 & 4.23 & 2.94 & 5.07  &   0.713 & 6118 & 5880 &  0.607  &6004 & 5855    \\
13683   &17 40 21.800 &53 43 56.60 & 15.776   &    5620 &  3.58 &  1.66 &  5.36 & 4.23 & 2.90 & 5.04  &   0.822 & 5747 & 5611 &  0.713  &5597 & 5494    \\
13885   &17 40 22.050 &53 37 54.10 & 15.906   &    5800 &  3.69 &  1.68 &  5.46 & 4.19 & 2.92 & 5.09  &   0.768 & 5922 & 5742 &  0.659  &5795 & 5672    \\
13909   &17 40 22.000 &53 35 22.20 & 15.956   &    5980 &  3.79 &  1.83 &  5.24 & 4.28 & 2.92 & 5.09  &   0.747 & 5995 & 5794 &  0.639  &5873 & 5741    \\
14522   &17 40 24.050 &53 44 26.50 & 16.207   &    6210 &  3.93 &  1.87 &  5.31 & 4.33 & 2.95 & 5.01  &   0.687 & 6216 & 5945 &  0.577  &6134 & 5966    \\
14589   &17 40 24.120 &53 42 36.40 & 16.089   &    6170 &  3.90 &  1.86 &  5.32 & 4.32 & 3.00 & 5.08  &   0.706 & 6145 & 5898 &  0.598  &6040 & 5885    \\
15344   &17 40 25.740 &53 35 20.90 & 15.890   &    5910 &  3.75 &  1.70 &  5.50 & 4.28 & 2.92 & 5.10  &   0.774 & 5899 & 5725 &  0.665  &5770 & 5650    \\
15957   &17 40 27.430 &53 36 53.80 & 16.287   &    6350 &  4.02 &  2.01 &  5.43 & 4.40 & 3.03 & 5.10  &   0.682 & 6238 & 5959 &  0.569  &6169 & 5995    \\
16588   &17 40 29.010 &53 37 42.30 & 15.837   &    5800 &  3.69 &  1.74 &  5.40 & 4.32 & 2.99 & 5.15  &   0.797 & 5826 & 5671 &  0.687  &5689 & 5578    \\
16792   &17 40 29.880 &53 44 49.70 & 15.956   &    5930 &  3.76 &  1.56 &  5.39 & 4.33 & 3.02 & 5.10  &   0.747 & 5995 & 5794 &  0.639  &5873 & 5741    \\
16822   &17 40 29.600 &53 37 38.40 & 15.860   &    5670 &  3.61 &  1.62 &  5.41 & 4.24 & 2.85 & 5.04  &   0.787 & 5857 & 5695 &  0.677  &5725 & 5610    \\
17100   &17 40 30.610 &53 43 52.70 & 16.002   &    6050 &  3.83 &  1.76 &  5.28 & 4.37 & 2.96 & 5.02  &   0.727 & 6064 & 5843 &  0.620  &5948 & 5806    \\
17167   &17 40 30.770 &53 44 01.20 & 16.014   &    6150 &  3.89 &  1.74 &  4.97 & 4.33 & 3.04 & 5.19  &   0.722 & 6083 & 5856 &  0.615  &5968 & 5823    \\
17629   &17 40 31.520 &53 37 03.30 & 15.779   &    5720 &  3.64 &  1.61 &  5.34 & 4.27 & 2.96 & 5.17  &   0.820 & 5750 & 5614 &  0.712  &5602 & 5498    \\
17841   &17 40 31.980 &53 36 38.00 & 15.764   &    5600 &  3.57 &  1.44 &  5.42 & 4.27 & 2.99 & 5.12  &   0.827 & 5731 & 5600 &  0.718  &5580 & 5478    \\
17925   &17 40 32.130 &53 35 56.30 & 16.295   &    6300 &  3.99 &  2.22 &  5.25 & 4.26 & 3.10 & 4.99  &   0.681 & 6240 & 5961 &  0.568  &6172 & 5998    \\
17964   &17 40 32.620 &53 43 41.70 & 16.037   &    6210 &  3.93 &  1.71 &  5.44 & 4.26 & 2.98 & 5.05  &   0.714 & 6115 & 5877 &  0.608  &5999 & 5851    \\
18670   &17 40 33.980 &53 37 14.90 & 16.297   &    6310 &  3.99 &  1.66 &  5.40 & 4.32 & 3.08 & 5.17  &   0.681 & 6240 & 5961 &  0.568  &6173 & 5999    \\
18866   &17 40 34.320 &53 35 54.20 & 15.888   &    5770 &  3.67 &  1.70 &  5.36 & 4.34 & 2.96 & 5.08  &   0.775 & 5896 & 5723 &  0.666  &5767 & 5647    \\
20071   &17 40 37.160 &53 37 29.50 & 15.551   &    5440 &  3.39 &  1.60 &  5.71 & 4.45 & 3.04 & 5.31  &   0.890 & 5547 & 5452 &  0.760  &5442 & 5351    \\
23267*  &17 40 43.750 &53 37 17.40 & 15.339   &    5430 &  3.36 &  1.60 &  5.46 & 4.38 & 2.98 & 5.25  &   0.920 & 5466 & 5385 &  0.778  &5386 & 5298    \\
2978    &17 39 37.960 &53 40 53.10 & 15.935   &    5810 &  3.70 &  1.70 &  5.24 & 4.13 & 2.89 & 5.00  &   0.755 & 5964 & 5772 &  0.647  &5840 & 5712    \\ 
3124    &17 39 38.870 &53 38 07.20 & 16.136   &    6190 &  3.91 &  1.88 &  5.31 & 4.33 & 3.00 & 5.04  &   0.698 & 6173 & 5917 &  0.590  &6077 & 5917    \\
\hline								       
\end{tabular}							       
\end{table}							       
\end{landscape}							       
\begin{landscape}						       
\begin{table}							       
\begin{tabular}{lllllllllllllllll}				       
\hline\hline 							       
ID  & $\alpha\rm(2000)$ & $\delta\rm(2000)$ &$V$& $T_{\rm H\alpha}$ & $\log{g}$ & $\xi$ & $\log{\epsilon_{\rm Mg}}$ & $\log{\epsilon_{\rm Ca}}$  & $\log{\epsilon_{\rm Ti}}$ & $\log{\epsilon_{\rm Fe}}$  & $v-y$ & $T_{v-y,\rm Dwarf}$ &$T_{v-y,\rm Giant}$ & $V-I$& $T_{V-I, \rm Dwarf}$& $T_{V-I, \rm Giant}$ \\ 
 &  & & & [K] &[cgs] & $\rm[km\,s^{-1}]$ & & & & & & [K]& [K] & & [K] & [K] \\
\hline								       
3296    &17 39 39.840 &53 37 07.30 & 15.619   &    5460 &  3.44 &  1.32 &  5.52 & 4.35 & 2.96 & 5.18  &   0.873 & 5596 & 5493 &  0.751  &5471 & 5377    \\
3602    &17 39 41.630 &53 34 56.90 & 15.981   &    5840 &  3.71 &  1.65 &  5.47 & 4.23 & 2.96 & 5.07  &   0.736 & 6032 & 5821 &  0.628  &5913 & 5776    \\
3796    &17 39 42.980 &53 39 25.30 & 16.231   &    6210 &  3.93 &  1.68 &  5.25 & 4.18 & 2.97 & 5.03  &   0.686 & 6223 & 5950 &  0.573  &6148 & 5978    \\
3838    &17 39 43.220 &53 38 59.40 & 16.292   &    6210 &  3.93 &  1.79 &  5.29 & 4.27 & 3.05 & 4.95  &   0.681 & 6239 & 5960 &  0.568  &6171 & 5997    \\
3867    &17 39 43.290 &53 35 09.50 & 15.743   &    5490 &  3.48 &  1.36 &  5.51 & 4.37 & 2.95 & 5.11  &   0.835 & 5705 & 5579 &  0.727  &5549 & 5450    \\
3892    &17 39 43.720 &53 40 50.60 & 15.966   &    5870 &  3.73 &  1.50 &  5.41 & 4.19 & 2.97 & 5.06  &   0.742 & 6010 & 5805 &  0.635  &5889 & 5755    \\
4114    &17 39 44.760 &53 39 46.50 & 16.263   &    6300 &  3.99 &  1.98 &  5.41 & 4.29 & 3.04 & 5.08  &   0.683 & 6231 & 5955 &  0.571  &6160 & 5988    \\
500199  &17 40 22.870 &53 31 56.40 & 15.621   &    5540 &  3.52 &  1.50 &  5.47 & 4.36 & 2.96 & 5.19  &   0.872 & 5597 & 5494 &  0.751  &5471 & 5378    \\
500645  &17 39 40.890 &53 32 55.80 & 16.264   &    6210 &  3.93 &  1.78 &  5.36 & 4.28 & 2.97 & 5.01  &   0.683 & 6232 & 5955 &  0.571  &6160 & 5988    \\
500725  &17 39 47.550 &53 33 03.70 & 16.013   &    5960 &  3.78 &  1.45 &  5.37 & 4.28 & 3.03 & 5.05  &   0.723 & 6081 & 5855 &  0.616  &5966 & 5822    \\
500949* &17 40 34.580 &53 33 20.70 & 15.514   &    5440 &  3.39 &  1.57 &  5.47 & 4.32 & 2.97 & 5.20  &   0.900 & 5520 & 5431 &  0.764  &5428 & 5338    \\
501164  &17 39 45.530 &53 33 48.60 & 15.906   &    6050 &  3.83 &  1.87 &  5.38 & 4.27 & 2.93 & 4.94  &   0.768 & 5922 & 5742 &  0.659  &5795 & 5672    \\
501236  &17 40 28.760 &53 33 55.70 & 16.213   &    6270 &  3.97 &  1.98 &  5.28 & 4.33 & 2.96 & 5.03  &   0.687 & 6218 & 5946 &  0.576  &6139 & 5970    \\
501262  &17 40 32.860 &53 33 58.40 & 16.137   &    6210 &  3.93 &  1.89 &  5.43 & 4.34 & 3.06 & 5.11  &   0.698 & 6174 & 5917 &  0.589  &6077 & 5918    \\
501368  &17 40 31.820 &53 34 14.30 & 16.180   &    6270 &  3.97 &  1.92 &  5.29 & 4.36 & 3.05 & 5.10  &   0.691 & 6199 & 5934 &  0.582  &6112 & 5947    \\
501389  &17 40 45.430 &53 34 17.50 & 16.197   &    6270 &  3.97 &  1.96 &  5.32 & 4.28 & 3.04 & 5.11  &   0.689 & 6210 & 5941 &  0.578  &6126 & 5959    \\
501939  &17 39 22.590 &53 35 52.70 & 16.049   &    6080 &  3.85 &  1.76 &  5.33 & 4.29 & 2.97 & 5.01  &   0.712 & 6122 & 5882 &  0.605  &6009 & 5859    \\
502286  &17 40 46.740 &53 36 49.10 & 16.016   &    6260 &  3.96 &  1.78 &  5.60 & 4.22 & 2.88 & 5.15  &   0.721 & 6086 & 5858 &  0.614  &5971 & 5826    \\
502729  &17 39 22.140 &53 38 23.40 & 15.831   &    5730 &  3.65 &  1.56 &  5.35 & 4.29 & 2.95 & 5.14  &   0.799 & 5818 & 5665 &  0.689  &5680 & 5569    \\
502740  &17 39 10.480 &53 38 25.90 & 15.607   &    5430 &  3.36 &  1.27 &  5.49 & 4.38 & 3.07 & 5.24  &   0.876 & 5587 & 5486 &  0.752  &5465 & 5373    \\
502882  &17 39 22.740 &53 38 51.90 & 15.801   &    5530 &  3.51 &  1.44 &  5.63 & 4.26 & 2.92 & 5.14  &   0.811 & 5779 & 5636 &  0.702  &5635 & 5528    \\
504920  &17 40 37.920 &53 44 02.80 & 15.898   &    5990 &  3.80 &  1.55 &  5.55 & 4.28 & 2.98 & 5.25  &   0.771 & 5910 & 5734 &  0.662  &5783 & 5661    \\
505031  &17 40 37.050 &53 44 17.70 & 15.770   &    5920 &  3.76 &  1.88 &  5.49 & 4.25 & 2.94 & 5.20  &   0.824 & 5739 & 5606 &  0.716  &5588 & 5486    \\
505253  &17 40 41.730 &53 44 40.90 & 15.923   &    5950 &  3.77 &  1.82 &  5.51 & 4.36 & 3.02 & 5.17  &   0.760 & 5946 & 5760 &  0.652  &5821 & 5695    \\
505471  &17 40 36.670 &53 45 02.80 & 16.122   &    6490 &  4.12 &  2.06 &  5.46 & 4.32 & 3.02 & 5.08  &   0.700 & 6165 & 5911 &  0.592  &6065 & 5908    \\
505513  &17 40 42.670 &53 45 05.60 & 16.226   &    6290 &  3.98 &  1.99 &  5.46 & 4.36 & 2.99 & 5.09  &   0.686 & 6221 & 5949 &  0.574  &6146 & 5976    \\
505913  &17 40 22.730 &53 45 37.30 & 15.928   &    5860 &  3.73 &  1.60 &  5.45 & 4.27 & 2.89 & 5.12  &   0.758 & 5953 & 5765 &  0.650  &5829 & 5702    \\
506120* &17 40 41.590 &53 45 49.30 & 16.271   &    6220 &  3.93 &  2.27 &  5.17 & 4.27 & 2.99 & 4.95  &   0.683 & 6233 & 5957 &  0.570  &6163 & 5990    \\
506139  &17 40 31.030 &53 45 51.80 & 15.834   &    5940 &  3.77 &  1.68 &  5.46 & 4.25 & 3.00 & 5.09  &   0.798 & 5822 & 5668 &  0.688  &5685 & 5573    \\
507010  &17 40 42.060 &53 46 45.70 & 15.248   &    5400 &  3.29 &  1.54 &  5.56 & 4.39 & 3.03 & 5.25  &   0.930 & 5441 & 5364 &  0.783  &5370 & 5284    \\
507031  &17 40 36.410 &53 46 47.70 & 16.048   &    6180 &  3.91 &  1.77 &  5.32 & 4.06 & 2.87 & 4.90  &   0.712 & 6121 & 5882 &  0.606  &6008 & 5858    \\
507433* &17 40 16.070 &53 47 18.60 & 16.278   &    6350 &  4.02 &  1.82 &  5.44 & 4.34 & 3.13 & 5.14  &   0.682 & 6235 & 5958 &  0.570  &6166 & 5993    \\
507860  &17 40 33.610 &53 47 52.40 & 16.238   &    6240 &  3.95 &  1.46 &  5.47 & 4.35 & 2.94 & 4.96  &   0.685 & 6225 & 5951 &  0.573  &6151 & 5980    \\
508219  &17 40 17.200 &53 48 24.40 & 16.285   &    6350 &  4.02 &  1.74 &  5.50 & 4.34 & 3.03 & 5.10  &   0.682 & 6237 & 5959 &  0.569  &6168 & 5995    \\
5734    &17 39 52.820 &53 37 03.00 & 16.148   &    6120 &  3.87 &  1.70 &  5.20 & 4.25 & 2.98 & 5.01  &   0.696 & 6180 & 5921 &  0.587  &6086 & 5925    \\
6114    &17 39 54.710 &53 37 40.20 & 15.947   &    5770 &  3.67 &  1.60 &  5.32 & 4.13 & 2.84 & 4.95  &   0.750 & 5981 & 5785 &  0.642  &5859 & 5728    \\
6119    &17 39 54.910 &53 42 24.20 & 16.171   &    6210 &  3.93 &  1.75 &  5.38 & 4.33 & 3.03 & 5.07  &   0.693 & 6194 & 5931 &  0.583  &6105 & 5941    \\
6391*   &17 39 55.900 &53 35 11.90 & 15.551   &    5530 &  3.51 &  1.61 &  5.47 & 4.39 & 3.03 & 5.23  &   0.890 & 5547 & 5452 &  0.760  &5442 & 5351    \\ 
6550    &17 39 56.790 &53 40 29.90 & 15.954   &    5990 &  3.80 &  1.78 &  5.39 & 4.24 & 2.98 & 5.04  &   0.747 & 5992 & 5792 &  0.639  &5870 & 5738    \\
6760    &17 39 57.710 &53 38 39.50 & 15.832   &    5710 &  3.64 &  1.53 &  5.32 & 4.27 & 2.95 & 5.10  &   0.799 & 5819 & 5666 &  0.689  &5682 & 5571    \\
6997    &17 39 58.900 &53 43 16.10 & 15.976   &    5970 &  3.79 &  1.69 &  5.36 & 4.24 & 2.94 & 5.07  &   0.738 & 6025 & 5815 &  0.630  &5905 & 5769    \\
7081    &17 39 59.050 &53 38 04.50 & 15.984   &    5860 &  3.73 &  1.72 &  5.37 & 4.27 & 2.90 & 4.99  &   0.735 & 6037 & 5824 &  0.627  &5918 & 5780    \\
7090    &17 39 59.350 &53 44 36.10 & 16.164   &    6100 &  3.86 &  1.61 &  5.41 & 4.16 & 2.97 & 5.03  &   0.694 & 6190 & 5928 &  0.585  &6099 & 5936    \\
7181    &17 39 59.450 &53 38 38.30 & 15.582   &    5490 &  3.48 &  1.46 &  5.65 & 4.43 & 3.04 & 5.28  &   0.882 & 5569 & 5471 &  0.756  &5455 & 5363    \\
7512    &17 40 00.780 &53 35 35.10 & 15.944   &    5950 &  3.77 &  1.92 &  5.30 & 4.19 & 2.96 & 5.05  &   0.752 & 5977 & 5782 &  0.643  &5854 & 5724    \\
7599    &17 40 01.210 &53 37 52.70 & 16.073   &    6170 &  3.90 &  1.86 &  5.33 & 4.29 & 2.94 & 5.06  &   0.708 & 6136 & 5892 &  0.601  &6027 & 5875    \\
7611    &17 40 01.520 &53 43 46.80 & 15.932   &    5770 &  3.67 &  1.59 &  5.42 & 4.23 & 2.94 & 5.06  &   0.757 & 5959 & 5769 &  0.648  &5835 & 5707    \\
7698    &17 40 01.580 &53 36 49.30 & 15.703   &    5620 &  3.58 &  1.61 &  5.54 & 4.31 & 2.99 & 5.21  &   0.851 & 5659 & 5543 &  0.740  &5506 & 5410    \\
7720    &17 40 01.820 &53 40 55.00 & 15.728   &    5580 &  3.55 &  1.39 &  5.74 & 4.56 & 3.20 & 5.35  &   0.841 & 5687 & 5565 &  0.733  &5528 & 5430    \\
\hline								       
\end{tabular}							       
\end{table}							       
\end{landscape}							       
\begin{landscape}						       
\begin{table}							       
\begin{tabular}{lllllllllllllllll}				       
\hline\hline 							       
ID  & $\alpha\rm(2000)$ & $\delta\rm(2000)$ &$V$& $T_{\rm H\alpha}$ & $\log{g}$ & $\xi$ & $\log{\epsilon_{\rm Mg}}$ & $\log{\epsilon_{\rm Ca}}$  & $\log{\epsilon_{\rm Ti}}$ & $\log{\epsilon_{\rm Fe}}$  & $v-y$ & $T_{v-y,\rm Dwarf}$ &$T_{v-y,\rm Giant}$ & $V-I$& $T_{V-I, \rm Dwarf}$& $T_{V-I, \rm Giant}$ \\ 
 &  & & & [K] &[cgs] & $\rm[km\,s^{-1}]$ & & & & & & [K]& [K] & & [K] & [K] \\
\hline								       
8075    &17 40 02.960 &53 35 35.60 & 15.923   &    5820 &  3.70 &  1.54 &  5.42 & 4.22 & 2.91 & 5.03  &   0.760 & 5946 & 5760 &  0.652  &5821 & 5695    \\
8087    &17 40 03.320 &53 42 46.80 & 16.229   &    6360 &  4.03 &  1.59 &  5.33 & 4.37 & 3.07 & 5.13  &   0.686 & 6222 & 5949 &  0.574  &6147 & 5977    \\
8099    &17 40 03.120 &53 36 46.90 & 15.651   &    5530 &  3.51 &  1.45 &  5.57 & 4.31 & 3.00 & 5.18  &   0.864 & 5620 & 5512 &  0.747  &5484 & 5390    \\
8171    &17 40 03.760 &53 45 09.40 & 16.243   &    6410 &  4.06 &  1.76 &  5.49 & 4.32 & 3.08 & 5.20  &   0.685 & 6226 & 5952 &  0.572  &6153 & 5982    \\
8320    &17 40 04.330 &53 43 05.80 & 16.027   &    6020 &  3.81 &  1.67 &  5.31 & 4.25 & 2.95 & 5.08  &   0.717 & 6103 & 5870 &  0.610  &5989 & 5842    \\
8344    &17 40 04.310 &53 39 30.30 & 15.971   &    5960 &  3.78 &  1.57 &  5.37 & 4.31 & 3.01 & 5.08  &   0.740 & 6017 & 5810 &  0.633  &5897 & 5762    \\
8395    &17 40 04.770 &53 45 30.60 & 15.981   &    5970 &  3.79 &  1.59 &  5.14 & 4.19 & 2.91 & 5.01  &   0.736 & 6032 & 5821 &  0.628  &5913 & 5776    \\
8511    &17 40 04.940 &53 40 24.70 & 15.653   &    5560 &  3.54 &  1.66 &  5.53 & 4.29 & 2.93 & 5.15  &   0.864 & 5621 & 5513 &  0.746  &5485 & 5391    \\
8598    &17 40 05.430 &53 44 10.30 & 16.160   &    6220 &  3.93 &  2.15 &  5.32 & 4.32 & 3.04 & 5.05  &   0.695 & 6187 & 5926 &  0.585  &6096 & 5933    \\
8661    &17 40 05.430 &53 38 19.80 & 16.119   &    6240 &  3.95 &  1.83 &  5.25 & 4.37 & 3.03 & 5.13  &   0.701 & 6163 & 5910 &  0.593  &6063 & 5906    \\
8802    &17 40 06.100 &53 42 24.40 & 15.951   &    5920 &  3.76 &  1.83 &  5.46 & 4.25 & 2.94 & 5.07  &   0.749 & 5987 & 5789 &  0.641  &5865 & 5734    \\
8848    &17 40 06.390 &53 45 36.00 & 16.032   &    5870 &  3.73 &  1.72 &  5.11 & 4.16 & 2.89 & 4.93  &   0.715 & 6111 & 5875 &  0.609  &5996 & 5847    \\
8870    &17 40 06.350 &53 42 09.00 & 15.973   &    5930 &  3.76 &  1.79 &  5.35 & 4.25 & 3.00 & 5.11  &   0.739 & 6020 & 5812 &  0.632  &5900 & 5765    \\
8901    &17 40 06.150 &53 34 36.50 & 15.820   &    5730 &  3.65 &  1.68 &  5.44 & 4.28 & 2.97 & 5.07  &   0.804 & 5803 & 5655 &  0.694  &5663 & 5554    \\
9173    &17 40 07.490 &53 44 47.50 & 16.117   &    6080 &  3.85 &  1.71 &  5.37 & 4.28 & 2.98 & 5.00  &   0.701 & 6162 & 5909 &  0.593  &6062 & 5904    \\
9226    &17 40 07.200 &53 34 59.90 & 16.225   &    6230 &  3.94 &  1.99 &  5.36 & 4.32 & 2.98 & 5.01  &   0.686 & 6221 & 5949 &  0.574  &6146 & 5976    \\
9461    &17 40 08.400 &53 42 51.60 & 15.934   &    5850 &  3.72 &  1.67 &  5.28 & 4.20 & 2.89 & 5.08  &   0.756 & 5962 & 5771 &  0.647  &5838 & 5710    \\
9483    &17 40 08.420 &53 41 35.10 & 15.568   &    5530 &  3.51 &  1.59 &  5.53 & 4.36 & 2.97 & 5.22  &   0.886 & 5559 & 5462 &  0.757  &5449 & 5358    \\
9655    &17 40 09.090 &53 43 26.40 & 16.200   &    6330 &  4.01 &  2.12 &  5.26 & 4.33 & 2.99 & 5.05  &   0.688 & 6212 & 5942 &  0.578  &6128 & 5961    \\
9909*   &17 40 09.530 &53 34 26.10 & 15.947   &    5910 &  3.75 &  1.63 &  5.40 & 4.26 & 3.02 & 5.03  &   0.750 & 5981 & 5785 &  0.642  &5859 & 5728    \\  
\hline
\end{tabular}
\end{table}
\end{landscape}

\begin{longtable}{lrrrcrrrrrr}
\caption{List of lines used in the abundance analysis, with references to the oscillator strengths used. The columns Waals.\,, Stark.\,, and Rad.\, list the van der Waals, Stark, and radiative damping data  (for a description of the data sources, see Barklem et al.\ 2005, section 3.2, paragraph 5). Rad.\ is the logarithm (base 10) of the FWHM given in $\rm rad\,s^{-1}$. Non-zero values of Stark.\, and negative values of Waals.\, represent the logarithm of FWHM per unit pertuber number density at 10\,000\,K, given in $\rm rad\,s^{-1}cm^{-3}$. Positive values of the van der Waals parameter correspond to the notation of \citet{Anstee95}, where a packed parameter is used for the broadening cross-section ($\sigma$) for collisions by neutral hydrogen at 10$\rm km\,s^{-1}$ and its velocity dependence ($\alpha$). E.g. for the Mg\,I 5172.684\AA\ line, Waals.$\,=729.238$ means $\sigma=729$ in atomic units and $\alpha=0.238$. d$\lambda_{\rm blue}$ and d$\lambda_{\rm red}$ give the extent of the spectral windows used for fitting of each line, blue-wards and red-wards of the central wavelength. Zero values in both columns indicate that the line-contribution is considered in the same window as the one listed directly above.}\\
\hline\hline
Ion & $\lambda$ [\AA] & $\chi$ [eV] & $\log{gf}$ &  Ref. &  Waals.  &  Stark.  &  Rad.    &   d$\lambda_{\rm blue}$ [m\AA]  & d$\lambda_{\rm red}$ [m\AA]\\
\hline
\endfirsthead
\caption{Continued.}\\
\hline\hline
Ion & $\lambda$ [\AA] & $\chi$ [eV] & $\log{gf}$ &  Ref. &  Waals.  &  Stark.  &  Rad.    &   d$\lambda_{\rm blue}$ [m\AA]  & d$\lambda_{\rm red}$ [m\AA]\\
\hline
\endhead
\hline
\endfoot
Mg\,I & 5172.684 &  2.712 & -0.380 & $^1$    & 729.238  &      0. & 7.990 & -700 &  700 \\ 
Mg\,I & 5183.604 &  2.717 & -0.160 & $^1$    & 729.238  &      0. & 7.990 & -700 &  700 \\ 
Ca\,I & 4434.957 &  1.886 & -0.007 & $^2$    &  -7.162  & -5.602  & 8.021 & -300 &  100 \\ 
Ca\,I & 4435.679 &  1.886 & -0.517 & $^2$    &  -7.163  & -5.610  & 8.025 & -300 &  300 \\ 
Ca\,I & 4454.779 &  1.899 &  0.258 & $^2$    &  -7.162  & -5.596  & 8.017 & -100 &  100 \\ 
Ca\,I & 4455.887 &  1.899 & -0.518 & $^2$    &  -7.162  & -5.602  & 8.021 & -300 &  300 \\ 
Ca\,I & 5261.704 &  2.521 & -0.579 & $^3$    &  -7.416  & -5.756  & 7.903 & -300 &  200 \\ 
Ca\,I & 5265.556 &  2.523 & -0.113 & $^3$    &  -7.416  & -5.755  & 7.903 & -300 &   50 \\ 
Ca\,I & 6439.075 &  2.526 &  0.390 & $^3$    &  -7.704  & -6.072  & 7.649 & -300 &  300 \\ 
Ca\,I & 6462.567 &  2.523 &  0.262 & $^3$    &  -7.704  & -6.072  & 7.645 & -300 &   50 \\ 
Ca\,I & 6493.781 &  2.521 & -0.109 & $^3$    &  -7.704  & -6.071  & 7.640 & -300 &  300 \\ 
Ti\,I & 4533.241 &  0.848 &  0.476 & $^4$    &  -7.593  & -5.144  & 8.083 & -300 &  300 \\ 
Ti\,I & 4534.776 &  0.836 &  0.280 & $^4$    &  -7.596  & -5.313  & 8.079 & -300 &  300 \\ 
Ti\,I & 4535.568 &  0.826 &  0.161 & $^5$    &  -7.840  & -5.403  & 8.079 & -300 &   50 \\ 
Ti\,II& 4394.059 &  1.221 & -1.780 & $^6$    &  -7.944  & -6.601  & 8.471 & -100 &  300 \\ 
Ti\,II& 4395.840 &  1.243 & -1.930 & $^6$    &  -7.904  & -6.550  & 8.471 & -200 &  200 \\ 
Ti\,II& 4399.765 &  1.237 & -1.190 & $^6$    &  -7.946  & -6.612  & 8.461 & -300 &  200 \\ 
Ti\,II& 4417.714 &  1.165 & -1.190 & $^6$    &  -7.926  & -6.665  & 8.225 & -300 &  300 \\ 
Ti\,II& 4443.801 &  1.080 & -0.720 & $^6$    &  -7.923  & -6.509  & 8.199 & -200 &  300 \\ 
Ti\,II& 4450.482 &  1.084 & -1.520 & $^6$    &  -7.920  & -6.502  & 8.199 & -300 &  200 \\ 
Ti\,II& 4468.507 &  1.131 & -0.600 & $^7$    &  -7.931  & -6.723  & 8.207 & -300 &  300 \\ 
Ti\,II& 4470.853 &  1.165 & -2.020 & $^6$    &  -7.928  & -6.713  & 8.217 & -250 &  150 \\ 
Ti\,II& 4501.270 &  1.116 & -0.770 & $^6$    &  -7.931  & -6.729  & 8.199 & -300 &  300 \\ 
Ti\,II& 4533.960 &  1.237 & -0.530 & $^6$    &  -7.960  & -6.661  & 8.225 & -300 &  100 \\ 
Ti\,II& 5185.902 &  1.893 & -1.490 & $^6$    &  -7.908  & -6.533  & 8.367 & -300 &  300 \\ 
Ti\,II& 5226.538 &  1.566 & -1.260 & $^6$    &  -7.953  & -6.713  & 8.217 & -300 &   70 \\ 
Fe\,I & 4375.930 &  0.000 & -3.031 & $^9$    & 215.249  & -6.320  & 4.622 & -150 &  300 \\ 
Fe\,I & 4375.986 &  3.047 & -2.029 & $^8$    &  -7.800  & -5.760  & 7.810 &    0 &    0 \\ 
Fe\,I & 4383.545 &  1.485 &  0.208 & $^9$    & 295.265  & -6.200  & 7.936 & -300 &  300 \\ 
Fe\,I & 4404.750 &  1.557 & -0.147 & $^9$    & 301.263  & -6.200  & 7.969 & -250 &  200 \\ 
Fe\,I & 4415.123 &  1.608 & -0.621 & $^9$    & 308.257  & -6.200  & 7.986 & -150 &  200 \\ 
Fe\,I & 4427.310 &  0.052 & -2.924 & $^9$    &  -7.880  & -6.320  & 4.696 & -100 &  300 \\ 
Fe\,I & 4430.614 &  2.223 & -1.728 & $^9$    & 431.302  & -6.080  & 8.606 & -250 &  250 \\ 
Fe\,I & 4442.339 &  2.198 & -1.228 & $^9$    & 424.302  & -6.080  & 8.599 & -300 &  300 \\ 
Fe\,I & 4443.194 &  2.858 & -1.043 & $^9$    & 224.263  & -6.340  & 7.799 & -150 &  250 \\ 
Fe\,I & 4443.196 &  3.071 & -1.905 & $^8$    &  -7.770  & -6.020  & 8.010 &    0 &    0 \\ 
Fe\,I & 4447.717 &  2.223 & -1.339 & $^9$    & 429.302  & -6.080  & 8.604 & -250 &  200 \\ 
Fe\,I & 4461.653 &  0.087 & -3.210 & $^7$    &  -7.799  & -6.320  & 4.638 & -200 &  200 \\ 
Fe\,I & 4476.019 &  2.845 & -0.819 & $^9$    &  -7.830  & -6.170  & 7.825 & -300 &  300 \\ 
Fe\,I & 4476.076 &  3.686 & -0.175 & $^8$    &  -7.670  & -4.730  & 7.935 &    0 &    0 \\ 
Fe\,I & 4482.170 &  0.110 & -3.501 & $^7$    &  -7.799  & -6.320  & 4.529 &    0 &    0 \\ 
Fe\,I & 4482.253 &  2.223 & -1.482 & $^9$    &  -7.788  & -6.080  & 8.599 &    0 &    0 \\ 
Fe\,I & 4489.739 &  0.121 & -3.899 & $^9$    & 218.249  & -6.320  & 4.403 & -300 &  200 \\ 
Fe\,I & 4494.563 &  2.198 & -1.143 & $^9$    & 416.302  & -6.080  & 8.600 & -300 &  300 \\ 
Fe\,I & 4528.614 &  2.176 & -0.887 & $^9$    & 407.301  & -6.080  & 8.607 & -100 &  300 \\ 
Fe\,I & 4531.148 &  1.485 & -2.155 & $^7$    &  -7.678  & -6.200  & 8.083 & -250 &  300 \\ 
Fe\,I & 4736.773 &  3.211 & -0.752 & $^9$    & 820.231  & -5.290  & 7.798 & -200 &  200 \\ 
Fe\,I & 4871.318 &  2.865 & -0.363 & $^9$    & 748.235  & -5.490  & 8.005 & -300 &  300 \\ 
Fe\,I & 4872.138 &  2.882 & -0.567 & $^9$    & 754.235  & -5.490  & 8.004 & -300 &  300 \\ 
Fe\,I & 4890.755 &  2.875 & -0.394 & $^9$    & 758.236  & -5.240  & 8.004 & -350 &  300 \\ 
Fe\,I & 4891.492 &  2.851 & -0.112 & $^9$    & 750.237  & -5.490  & 8.009 & -300 &  350 \\ 
Fe\,I & 4903.310 &  2.882 & -0.926 & $^9$    &  -7.259  & -5.490  & 8.004 & -300 &  300 \\ 
Fe\,I & 4918.994 &  2.865 & -0.342 & $^9$    & 750.237  & -5.490  & 8.009 & -350 &  350 \\ 
Fe\,I & 4918.954 &  4.154 & -0.635 & $^8$    &  -7.510  & -4.690  & 8.470 &    0 &    0 \\ 
Fe\,I & 4920.503 &  2.832 &  0.068 & $^9$    & 739.238  & -5.490  & 8.009 & -300 &  300 \\ 
Fe\,I & 5150.840 &  0.990 & -3.037 & $^9$    &  -7.742  & -6.250  & 7.180 & -300 &  300 \\ 
Fe\,I & 5166.282 &  0.000 & -4.123 & $^9$    &  -7.826  & -6.330  & 3.540 & -180 &  180 \\ 
Fe\,I & 5171.596 &  1.485 & -1.721 & $^9$    &  -7.687  & -6.200  & 6.330 & -220 &  220 \\ 
Fe\,I & 5191.455 &  3.038 & -0.551 & $^9$    &  -7.258  & -5.490  & 8.004 & -300 &  300 \\ 
Fe\,I & 5192.344 &  2.998 & -0.421 & $^9$    &  -7.266  & -5.490  & 8.010 & -300 &  250 \\ 
Fe\,I & 5194.942 &  1.557 & -2.021 & $^9$    &  -7.680  & -6.200  & 6.290 & -300 &  200 \\ 
Fe\,I & 5202.336 &  2.176 & -1.838 & $^7$    &  -7.603  & -6.180  & 8.230 & -250 &  250 \\ 
Fe\,I & 5202.256 &  4.256 & -0.837 & $^8$    &  -7.765  & -6.010  & 8.320 &    0 &    0 \\ 
Fe\,I & 5216.274 &  1.608 & -2.082 & $^9$    &  -7.674  & -6.200  & 6.220 & -300 &  300 \\ 
Fe\,I & 5232.940 &  2.940 & -0.058 & $^9$    &  -7.280  & -5.490  & 8.009 & -300 &  300 \\ 
Fe\,I & 5266.554 &  2.998 & -0.386 & $^9$    &  -7.273  & -5.489  & 8.009 & -300 &  300 \\ 
Fe\,I & 5269.537 &  0.859 & -1.321 & $^7$    &  -7.761  & -6.300  & 7.185 & -350 &  300 \\ 
Fe\,I & 5281.790 &  3.038 & -0.834 & $^9$    &  -7.266  & -5.490  & 8.010 & -200 &  200 \\ 
Fe\,I & 5283.621 &  3.241 & -0.524 & $^9$    &  -7.221  & -5.450  & 7.880 &  -50 &  250 \\ 
Fe\,I & 5324.191 &  3.211 & -0.103 & $^{10}$ &  -7.235  & -5.500  & 7.880 & -250 &  250 \\ 
Fe\,I & 5328.039 &  0.915 & -1.466 & $^7$    &  -7.757  & -6.302  & 7.161 & -300 &  150 \\ 
Fe\,I & 5328.532 &  1.557 & -1.850 & $^9$    &  -7.686  & -6.228  & 6.848 & -150 &  300 \\ 
Fe\,I & 5339.929 &  3.266 & -0.647 & $^{10}$ &  -7.221  & -5.451  & 7.874 & -250 &  250 \\ 
Fe\,I & 6393.601 &  2.433 & -1.576 & $^9$    &  -7.622  & -6.310  & 7.970 & -250 &  250 \\ 
Fe\,I & 6400.001 &  3.602 & -0.290 & $^{10}$ &  -7.232  & -5.500  & 7.900 & -250 &  200 \\ 
Fe\,I & 6430.846 &  2.176 & -1.946 & $^9$    &  -7.704  & -6.190  & 8.220 & -220 &  220 \\ 
Fe\,I & 6494.981 &  2.404 & -1.273 & $^7$    &  -7.629  & -6.330  & 7.936 & -300 &  300 \\ 
Fe\,II& 4416.828 &  2.778 & -2.540 & $^{11}$ &  -7.950  & -6.670  & 8.614 & -200 &  300 \\ 
Fe\,II& 4491.405 &  2.856 & -2.700 & $^{12}$ &  -7.950  & -6.600  & 8.481 & -200 &  200 \\ 
Fe\,II& 4508.288 &  2.856 & -2.312 & $^{13}$ &  -7.950  & -6.670  & 8.617 & -250 &  250 \\ 
Fe\,II& 4515.343 &  2.844 & -2.362 & $^{14}$ &  -7.950  & -6.600  & 8.487 & -200 &  200 \\ 
Fe\,II& 4923.927 &  2.891 & -1.206 & $^{14}$ &  -7.914  & -6.583  & 8.489 & -300 &  300 \\ 
Fe\,II& 5197.577 &  3.230 & -2.100 & $^{12}$ &  -7.824  & -6.600  & 8.480 & -200 &  200 \\ 
Fe\,II& 5234.625 &  3.221 & -2.270 & $^{15}$ &  -7.946  & -6.600  & 8.490 & -200 &  200 \\ 
Fe\,II& 5284.109 &  2.891 & -3.130 & $^{11}$ &  -7.914  & -6.600  & 8.530 & -160 &  160 \\ 
Fe\,II& 5316.615 &  3.153 & -1.850 & $^{12}$ &  -7.950  & -6.600  & 8.480 & -300 &  100 \\ 
\hline                                                                                             
\multicolumn{10}{l}{  }\\                                                                         
\multicolumn{10}{l}{$^{1}$  Wiese \& Martin (1980)}                    \\                            	
\multicolumn{10}{l}{$^{2}$  Smith \& O'Neill (1975)}                 \\                            	
\multicolumn{10}{l}{$^{3}$  Smith \& Raggett (1981)}                 \\                            
\multicolumn{10}{l}{$^{4}$  Blackwell et al (1982)}                  \\     
\multicolumn{10}{l}{$^{5}$  R.\,Kurucz, http://cfaku5.cfa.harvard.edu/ATOMS/2200} \\      
\multicolumn{10}{l}{$^{6}$  Pickering et al (2001)}                 \\                            
\multicolumn{10}{l}{$^{7}$  Martin et al. (1988)}                    \\                            
\multicolumn{10}{l}{$^{8}$  R.\,Kurucz, http://cfaku5.cfa.harvard.edu/ATOMS/2600} \\           
\multicolumn{10}{l}{$^{9}$  O'Brian et al. (1991)}  \\   %
\multicolumn{10}{l}{$^{10}$ Bard et al. (1991)}    \\   %
\multicolumn{10}{l}{$^{11}$ Moity (1983)}          \\   %
\multicolumn{10}{l}{$^{12}$ Kroll \& Kock (1987)}  \\   %
\multicolumn{10}{l}{$^{13}$ Biemont et al. (1991)} \\   %
\multicolumn{10}{l}{$^{14}$ Schnabel et al (2004)}       \\   %
\multicolumn{10}{l}{$^{15}$ Heise \& Kock (1990)}  \\   %
\multicolumn{10}{l}{  } \\    
\end{longtable}	      
\nocite{Martin88}
\nocite{Wiese80}   
\nocite{Obrian91}
\nocite{Smith75}    
\nocite{Blackwell82} 
\nocite{Bard91}
\nocite{Smith81}    
\nocite{Moity83}
\nocite{Pickering01}
\nocite{Kroll87}
\nocite{Biemont91}
\nocite{Schnabel04}
\nocite{Heise90}

\end{document}